\documentclass[twocolumn,superscriptaddress, aps,pre]{revtex4-1}
\makeatletter
\newcommand*{\rom}[1]{\expandafter\@slowromancap\romannumeral #1@}
\makeatother
\usepackage{color}
\usepackage{graphicx}
\usepackage{hyperref}
\usepackage{amsmath,amssymb}
\usepackage{epstopdf}
\usepackage{subcaption}
\graphicspath{{figs/}}
\def\beq{\begin{equation}}
\def\eeq{\end{equation}}
\def\bea{\begin{eqnarray}}
\def\eea{\end{eqnarray}}

\begin{document}

\title{Thermodynamic geometry and complexity of black holes in theories with broken translational invariance}
\author {H. Babaei-Aghbolagh}
\email{h.babaei@uma.ac.ir}
\affiliation{Department of Physics, University of Mohaghegh Ardabili, P.O. Box 179, Ardabil, Iran}
\author {Hosein Mohammadzadeh}
\email{mohammadzadeh@uma.ac.ir}
\affiliation{Department of Physics, University of Mohaghegh Ardabili, P.O. Box 179, Ardabil, Iran}
\author {Davood Mahdavian Yekta}
\email{d.mahdavian@hsu.ac.ir}
\affiliation{Department of Physics, Hakim Sabzevari University, P.O. Box 397, Sabzevar, Iran}
\author {Komeil Babaei Velni}
\email{babaeivelni@guilan.ac.ir}
\affiliation{Department of Physics, University of Guilan, P.O. Box 41335-1914, Rasht, Iran}
\affiliation{School of Physics, Institute for Research in Fundamental Sciences (IPM), P.O. Box 19395-5531,
Tehran, Iran}

\begin{abstract}
The relationship between thermodynamics and the Lloyd bound on the holographic complexity for a black hole has been of interest. We consider $D$ dimensional anti-de Sitter black holes with hyperbolic geometry as well as black holes with momentum relaxation that have a minimum for temperature and mass. We show that the singular points of the thermodynamic curvature of the black holes, as  thermodynamic systems, correspond to the zero points of the action and volume complexity at the Lloyd bound. For such black holes with a single horizon, the complexity of volume and the complexity of action at minimum mass and minimum temperature are zero, respectively.  We show that the thermodynamic curvature diverges at these minimal values. Because of the behaviour of action complexity and thermodynamic curvature at minimum temperature, we propose the action complexity as an order parameter of the black holes as thermodynamic systems. Also, we derive the critical exponent related to the thermodynamic curvature in different dimensions.
\end{abstract}
\maketitle

\section{Introduction}\label{1}
Following Hawking and Bekenstein outstanding papers, the black holes are considered as the thermodynamic systems \cite{hawking1976black,bardeen1973four,bekenstein2020generalized}. The thermodynamic laws about the black holes have been investigated for many different black holes. Also, some thermodynamic response functions for black holes are derived and according to their behaviours, some phase transition points are predicted \cite{davies1977thermodynamic,mansoori2014correspondence,banerjee2011thermodynamics,kol2006phase}. Recently, the thermodynamic geometry of many different black holes has been vastly investigated and some useful information is extracted. 

Thermodynamic geometry introduced by Ruppeiner and Weinhold based on the fluctuation theory \cite{ruppeiner1979thermodynamics,weinhold1975metric,ruppeiner1995riemannian}. For a thermodynamic system, one can construct a Riemann manifold using fluctuating thermodynamic parameters. Different metrics are defined for the thermodynamic parameters space. In entropy representation, the second derivatives of entropy with respect to related extensive parameters of the system such as internal energy, the volume of the system, and the total number of particles define the components of metric tensor \cite{ruppeiner1995riemannian}. Also, Weinhold introduced another metric which is based on the second derivatives of internal energy with respect to the related extensive parameters \cite{weinhold1975metric}. Of course, some different metrics can be found in the other representations \cite{janyszek1990riemannian}. The curvature of the constructed thermodynamic parameters space is the Ricci scalar of the thermodynamic manifold and has some information about the thermodynamic systems \cite{ruppeiner1995riemannian}.

It has been shown that the thermodynamic curvature of an ideal classical gas vanishes at all physical ranges \cite{salamon1984relation}. However, the thermodynamic curvature of an ideal Bose (Fermi) gas is positive (negative). The intrinsic statistical interaction of bosons is attractive while it is negative for fermions \cite{janyszek1990riemannian}. In fact, one can argue that the sign of thermodynamic curvature classifies the statistical interactions of the system. Also, the singular points of thermodynamic curvature can be evidence of the existence of phase transitions. For example, the thermodynamic geometry of ideal gases with particles that obey Bose-Einstein, $q$-deformed, Polychronakos, and non-extensive are singular at condensation transition point \cite{mirza2011condensation,mirza2010thermodynamic,mirza2011thermodynamic,adli2019nonperturbative,adli2019condensation}.

Thermodynamic geometry of well-known black holes such as Kerr, Kerr-Newman, Reissner–Nordstrom, BTZ, and so on has been investigated \cite{aaman2003geometry,ruppeiner2008thermodynamic,sahay2010thermodynamic,mirza2007ruppeiner,cai1999thermodynamic,sarkar2006thermodynamic,quevedo2009geometric}. Also, the thermodynamic geometry of some black holes in extended thermodynamic phase space has been considered \cite{zhang2015phase,gunasekaran2012extended,wang2022thermodynamic,bairagya2021geometry,mohammadzadeh2021thermodynamic}. It has been shown that the cosmological constant should be included to the fluctuating thermodynamics parameters to obtain a consistent corresponding Smarr relation for some black holes \cite{kubizvnak2017black,kastor2009enthalpy}. Also, for Born-Infeld black holes, the maximal field strength is included in fluctuating parameters \cite{bronnikov2017dyonic,yi2010energy}. We will consider the new parameter which is related to the momentum relaxation as a fluctuating parameter and extend the thermodynamic parameters space properly. We will obtain the thermodynamic curvature and consider its singular points. One can find some correspondence between the singular points of some response function and also, the complexity. 

Recently, the study of some concepts in quantum information theory such as  entanglement entropy \cite{ryu2006holographic}  and holographic complexity \cite{alishahiha2015holographic,brown2016holographic} has been considered using the approach of $\mathrm{AdS/CFT}$ correspondence. In fact, since the entanglement entropy is not enough to describe  the dynamic  of the black hole  beyond the horizon, another concept called the holographic complexity was introduced. According to \cite{brown2016holographic}, the minimum number of operators required to go from an initial state to a final state is defined as the complexity of the system. In the context of $\mathrm{AdS/CFT}$ correspondence \cite{maldacena1999large}, two prescriptions have been suggested for the holographic complexity; the “complexity=volume” (CV) conjecture \cite{susskind2016computational,stanford2014complexity} and the “complexity= action” (CA) conjecture \cite{brown2016holographic,carmi2017time}. In \cite{yekta2021holographic} we have employed the (CA) conjecture to investigate the action growth rate for charged and neutral $\mathrm{AdS}$ black branes of a holographic consisting of Einstein-Maxwell theory in $D$ dimensional bulk spacetime with $D-2$ massless scalar fields which is called Einstein-Maxwell-Axion (EMA) theory. In Ref. \cite{babaei2021complexity}, we have also found the complexity and its time evolution for charged $\mathrm{AdS}$ black holes (the Gubser-Rocha model \cite{gubser2010peculiar,gouteraux2014charge}) in diverse dimensions via the (CA) conjecture. It has been shown that the growth rate of the holographic complexity violates the Lloyd’s bound in early times. 

We will organize the paper as follows: black hole with hyperbolic geometry and its related thermodynamic geometry is investigated in Sec. \ref{2}. Also, the complexity of these black holes is considered using both action and volume conjectures in this section. The thermodynamic curvature of black holes with momentum relaxation is worked out in two thermodynamic phase spaces in Sec \ref{3}. Also, the action complexity, as well as the volume complexity, is computed. We show that one can find a meaningful relationship between the thermodynamic curvature and the complexity in some special points. In fact, the zero points of complexities correspond to the singular points of the thermodynamic curvature. Also, the thermodynamic curvature as a power law function is considered in the vicinity of critical points. We will conclude the paper in Sec. \ref{4}.
\section{$\mathrm{AdS}$ black holes with hyperbolic geometry}\label{2}
Black holes with hyperbolic horizons are well known to exist in $\mathrm{AdS}$ spacetime. Most of the black holes such as  Schwarzschild, Kerr and Kerr-Newman have an appropriate $\mathrm{AdS}$ analogues. Also, there exist some $\mathrm{AdS}$ black holes with no Ricci-flat analogous or the so called topological black holes that their horizons may either be spherical, planar or hyperbolic \cite{lemos1995three,lemos1996rotating,mann1997pair,vanzo1997black,chen2015deformed}. We briefly review the $\mathrm{AdS}$ black holes with hyperbolic geometry in this section. When a cosmological constant $\Lambda$  is included in the lagrangian, the Einstein-Hilbert action is changed to following form
\begin{equation}
S=\frac{1}{16\pi }\int \!  \sqrt{-g} \,\bigg( R-2\Lambda \bigg) \,d^Dx ,
\label{eq:action}
\end{equation}
where, $g=\det(g_{\mu\nu})$ is the determinant of the metric tensor, $R$ is the Ricci scalar, $D$ denotes the dimension of spacetime and $\Lambda$ is
\begin{equation}
\Lambda=-\frac{(D-1)(D-2)}{2L^2},
\label{eq:cosmocon}
\end{equation}
where $L$ is the $\mathrm{AdS}$ radius. In particular, we focus on the $\mathrm{AdS}$ black holes in $D$ dimensions, whose metric takes the general form
\begin{equation}\label{HigherDMetric}
d s^{2} = - f(r)\, d t^{2} + \frac{d r^{2}}{f(r)} + r^{2}\, d \Sigma^{2}_{k,D-2}\,,
\end{equation}
with
\begin{equation}\label{BlackeningFactor}
f(r) = \frac{r^2}{L^2}+k -
\frac{m_0}{r^{D-3}},
\end{equation}
where $k=\lbrace+1,0,-1\rbrace$ indicates the curvature of the ($D$--2)-dimensional line element $d \Sigma^{2}_{k,D-2}$,
which is given by
\begin{equation}
d\Sigma^2_{k,D-2}=d\theta^2+\sinh^2\theta\, d\Omega^2_{D-3} \  {\rm for\ }k=-1\,,
 \label{geometries}
\end{equation}
and $m_{0}$ is the mass parameter which is defined in the following. Also, with   $k=-1$, $d\Sigma^2_{-1,D-2}$ is the metric on a ($D$--2)-dimensional
hyperbolic `plane' with unit curvature. In particular, the black
holes corresponding to $k=\{+1, 0, -1\}$ have spherical, planar, and
hyperbolic horizons, respectively.
\subsection{Thermodynamic of black holes with hyperbolic geometry}
\begin{figure}
	\centerline{\includegraphics[scale=.40]{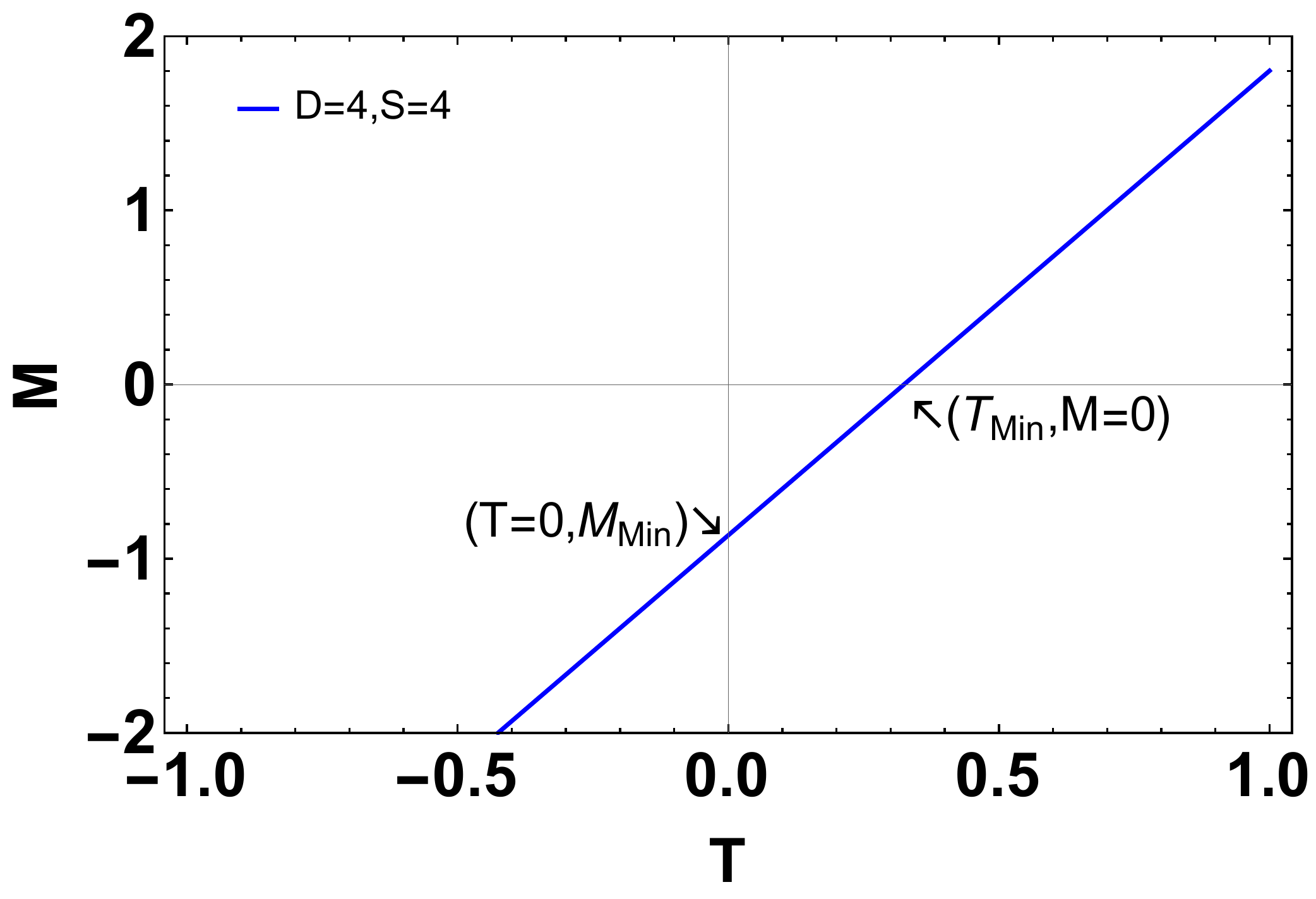}}
	\caption{Mass as a function of temperature for a fixed value of entropy in $D=4$ dimensional hyperbolic geometry.  }\label{fig:1}
\end{figure}

The location of the horizon $r_h$ is determined in terms of the `mass' parameter $m_0$ as follows
\begin{eqnarray}\label{eq-MQ}
 m_0&=& r_h^{D-3}\left(\frac{r_h^2}{L^2}-1\right),
\end{eqnarray}
and the mass is given by
\begin{eqnarray}\label{Mass}
M&=&\frac{(D-2)\mathcal{V}}{16\pi}m_0 \,,
\end{eqnarray}
where $\mathcal{V}$ denotes the dimensionless volume of the hyperbolic geometry in $D-2$ dimensions. The temperature of the thermal ensemble is given by
\beq
T=\frac{1}{4\pi}\left.\frac{\partial f}{\partial r}\right|_{r=r_h}=\frac{1}{4\pi r_h}\left((D-1)\,\frac{r_h^2}{L^2}- (D-3) \right)\,.
\label{temp}
\eeq
 Examining $f(r)$ in Eq. (\ref{BlackeningFactor}) with $k=-1$, we see that there is still a horizon
at $ r_h =  L$ even when \,$m_0=  0$.   Furthermore,  Eq. (\ref{temp}) then yields a finite temperature in this case as follows
\beq\label{tempmin}
T_{min}= \frac{1}{2\pi L}\,.
\eeq
In the context of $\mathrm{AdS/CFT}$, vacuum metric $M = 0$ has the form of an $\mathrm{AdS}$ black hole  that  can be interpreted in terms of an entangled state of two copies of the $\mathrm{CFT}$ on a hyperbolic plane
The entropy is related to the area of the horizon as follows
\begin{equation}\label{eq-S}
S=\frac{\mathcal{V}}{4}r_h^{D-2},
\end{equation}
and the thermodynamic pressure is defined in \cite{dolan2011pressure, cvetivc2011black, kubizvnak2012p} for the $\mathrm{AdS}$ spacetime 
\begin{equation}\label{eq-Pl}
P=-\frac{\Lambda}{8\pi}=\frac{(D-1)(D-2)}{16\pi L^2}.
\end{equation}
\begin{widetext}
Using Eqs. (\ref{eq-MQ}) and (\ref{temp}), and replacing $r_{h}$ and $L$ with entropy and pressure, we obtain the mass and temperature in terms of thermodynamic parameters as
\begin{eqnarray}\label{M1}
M&=&- \mathcal{V}^{\frac{1}{ D-2}} \frac{2^{-2-\frac{2}{ D-2}} ( D-2)  }{\pi} S^{ \frac{ D-3}{ D-2}} +\mathcal{V}^{\frac{1}{2 -  D}}  \frac{2^{2 + \frac{2}{D-2}} }{ D-1}P\,S^{1 + \frac{1}{ D-2}},
\end{eqnarray}
\begin{eqnarray}\label{T1}
T&=&- \mathcal{V}^{\frac{1}{ D-2}} \frac{2^{-2 -\frac{2}{D-2}} ( D-3) }{\pi}S^{\frac{1}{2 -  D}}+ \mathcal{V}^{\frac{1}{2 -  D}} \frac{2^{2 + \frac{2}{D-2}} }{D-2}P\,S^{\frac{1}{D-2}}.
\end{eqnarray}
\end{widetext}
Using Eq. (\ref{T1}), we can extract the pressure as a function of $S$ and $T$ and replace it in Eq. (\ref{M1}) to obtain an equation for $M$ as a function of the entropy and temperature. We have plotted $M$ as a function of $T$ for a fixed value of entropy in Fig. (\ref{fig:1}). One can define a minimum value for temperature; $T_{min}$, which corresponds to the $M=0$. In fact, for $T<T_{min}$ the mass of the black hole will be negative. Also, we can introduce a minimum mass; $M_{min}$, which corresponds to $T=0$. For the cases $M<M_{min}$, the temperature is negative. We work out the minimum temperature and minimum mass
\begin{eqnarray}
T_{min}&=& \mathcal{V}^{\frac{1}{ D-2}}  \frac{2^{-\frac{D}{ D-2}}}{\pi}  S^{-\frac{1}{ D-2}},
\end{eqnarray}
\begin{eqnarray}
M_{min}&=&- \mathcal{V}^{\frac{1}{ D-2}}\frac{2^{-\frac{D}{  D-2}} ( D-2)}{( D-1) \pi} S^{ \frac{ D-3}{ D-2}}.
\end{eqnarray}
We will explore some singular behaviour at these special points. In the following we consider the thermodynamic geometry and the thermodynamic curvature of thermodynamic parameters space. 

\subsection{Thermodynamic geometry}
We consider the thermodynamic geometry of $\mathrm{AdS}$ black holes in $D$ dimensional hyperbolic geometry. In order to construct the thermodynamic geometry, we select the appropriate fluctuating thermodynamic parameters. The Weinhold geometry is introduced by metric tensor components which are defined by the second derivative of internal energy with respect to the fluctuating parameters 
\begin{eqnarray}\label{weinhold}
  g_{ab} ^W =\frac{\partial^2 M}{\partial X^a \partial X^b}\,.
\end{eqnarray}
We note that the mass has the role of internal energy and the fluctuating parameters are considered  $X^1=S$  and $X^2=P$.
We will investigate the thermodynamic curvature using the Ruppeiner geometry\cite{hendi2015new,hendi2015geometrical,hendi2015phase}. We define the metric tensor components of Ruppeiner geometry 

\begin{eqnarray}\label{Ruppeiner}
ds^2_{R}=-M \,T^{-1}\,  g_{ab} ^W \,dX^a dX^b.
\end{eqnarray}
For a two dimensional thermodynamic parameters space, we evaluate the metric elements using Eqs. (\ref{M1}), (\ref{T1}), (\ref{weinhold}) and (\ref{Ruppeiner}).
 Using the metric tensor components, we obtain the affine connections, Riemann tensor, Ricci tensor, and finally the Ricci scalar of thermodynamic parameters space. The Ricci scalar or equivalently the thermodynamic curvature is given by:
 \begin{eqnarray}\label{R1}
\mathcal{R}&=& \pi {\mathcal{V}}^{\frac{7}{ D-2}}\frac{2^{3 + \frac{2}{D-2}} ( D-3) ( D-2)^2( D^2-1)}{X\,\, Y^3}  S^{ \frac{1}{ D-2}-2}\nonumber\\
&&+\pi^2 \mathcal{V}^{\frac{5}{ D-2}}\frac{2^{8 + \frac{6}{D-2}} ( D-3) (D-2) ( D-1) }{X\,\, Y^3} P S^{ \frac{3}{ D-2}-2}\nonumber\\
&&-3 \pi^3 \mathcal{V}^{\frac{3}{-2 + D}}\frac{2^{11 + \frac{10}{D-2}} ( D-1) }{X\,\,Y^3}P^2 S^{ \frac{5}{-2 + D}-2},
\end{eqnarray}
where,

	\begin{subequations}
	\begin{align}\label{XY}
	X=\pi\, 2^{4 + \frac{4}{ D-2}} P  S^{\frac{2}{ D-2}} - \mathcal{V}^{\frac{2}{ D-2}}  ( D-3) ( D-2)\, ,     
	\end{align}
	\begin{align} 
	Y=\pi 2^{4 + \frac{4}{ D-2}} P  S^{\frac{2}{ D-2}} - \mathcal{V}^{\frac{2}{ D-2}} ( D-2) ( D-1).
	\end{align}
\end{subequations}
It is obvious that the $X=0$ and $Y=0$ denote the singular points of the thermodynamic curvature. Using the above equations, we obtain spacial pressures $P_T$ ($X=0$) and $P_M$ ($Y=0$) 

\begin{subequations}
	\begin{align}\label{pt}
	P_T=\frac{\mathcal{V}^{\frac{2}{ D-2}}}{\pi}2^{-4 - \frac{4}{ D-2}} ( D-2) ( D-3) S^{-\frac{2}{ D-2}}\, ,     
	\end{align}
	\begin{align} 
	P_M=\frac{\mathcal{V}^{\frac{2}{ D-2}}}{\pi}2^{-4 - \frac{4}{ D-2}} ( D-1) ( D-2) S^{-\frac{2}{ D-2}}\nonumber\label{pm}.
	\end{align}
\end{subequations}
Using Eq. (\ref{T1}), One can extract the pressure as a function of entropy and temperature. Thus, we convert the Eq. (\ref{R1}) to a function of entropy and temperature as follows:
\begin{widetext}
 \begin{eqnarray}\label{RT}
\mathcal{R}&=&\frac{( D-1)}{( D-2)^2}\bigg(\frac{\mathcal{V}^{\frac{3}{ D-2}} 8^{\frac{D}{ D-2}} ( D-3) S^{-2 -  \frac{3}{ D-2}}}{ \pi^3 T (T -  T_{min})^3} -\frac{ \mathcal{V}^{\frac{2}{ D-2}}2^{\frac{ D+2}{ D-2}} ( D-3) S^{-2 -  \frac{2}{ D-2}} }{ \pi^2 (T -  T_{min})^3}-\frac{3 \mathcal{V}^{\frac{1}{ D-2}} 2^{\frac{D}{2 -  D}} S^{-2 -  \frac{1}{ D-2}} }{ \pi (T -  T_{min})^3}T \bigg)\,.
\end{eqnarray}

\begin{figure}[h]
	\begin{subfigure}{0.45\textwidth}\includegraphics[width=\textwidth]{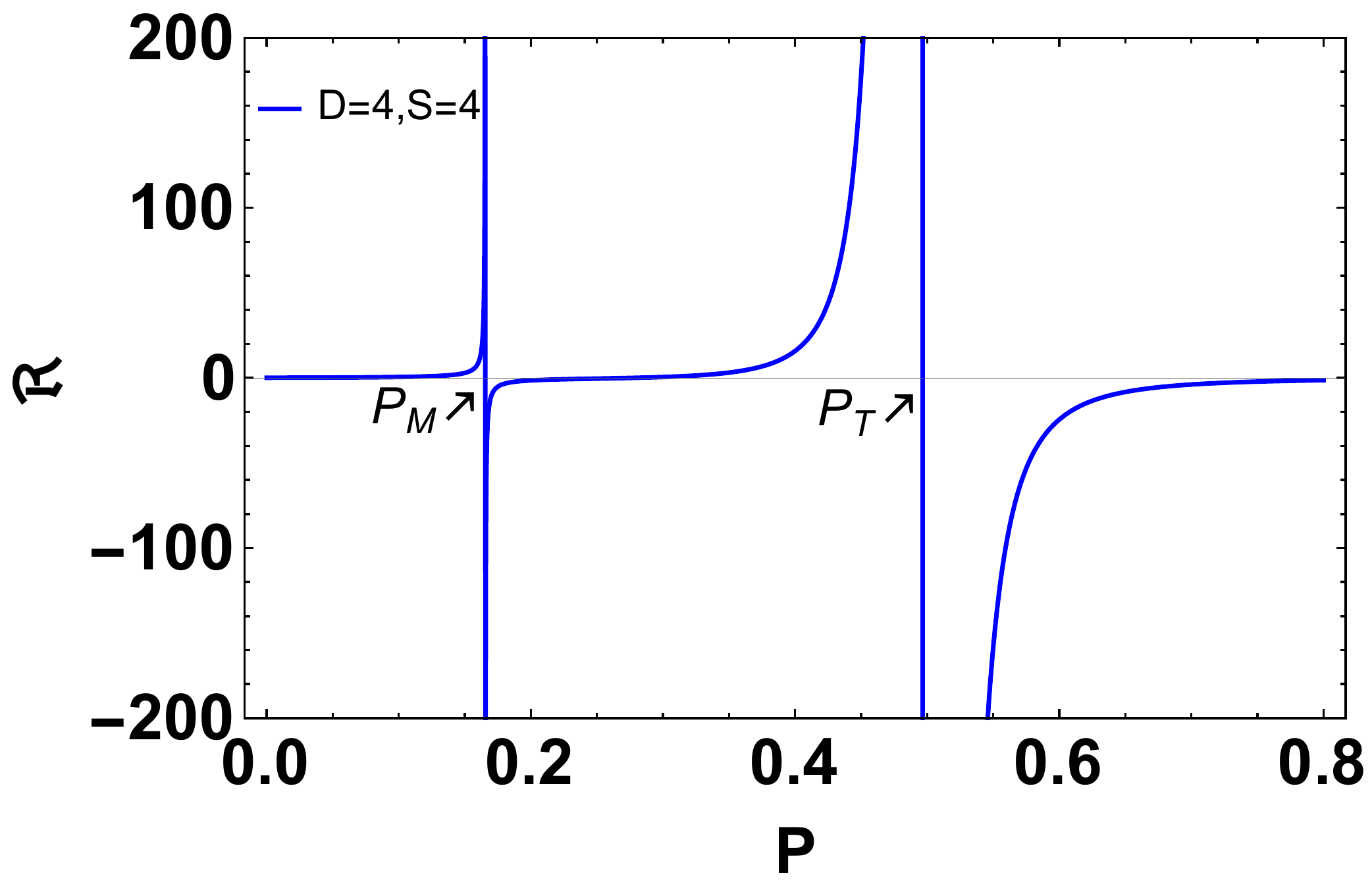}
		\caption{}
		\label{fig:2-1}
	\end{subfigure}
	\begin{subfigure}{0.45\textwidth}\includegraphics[width=\textwidth]{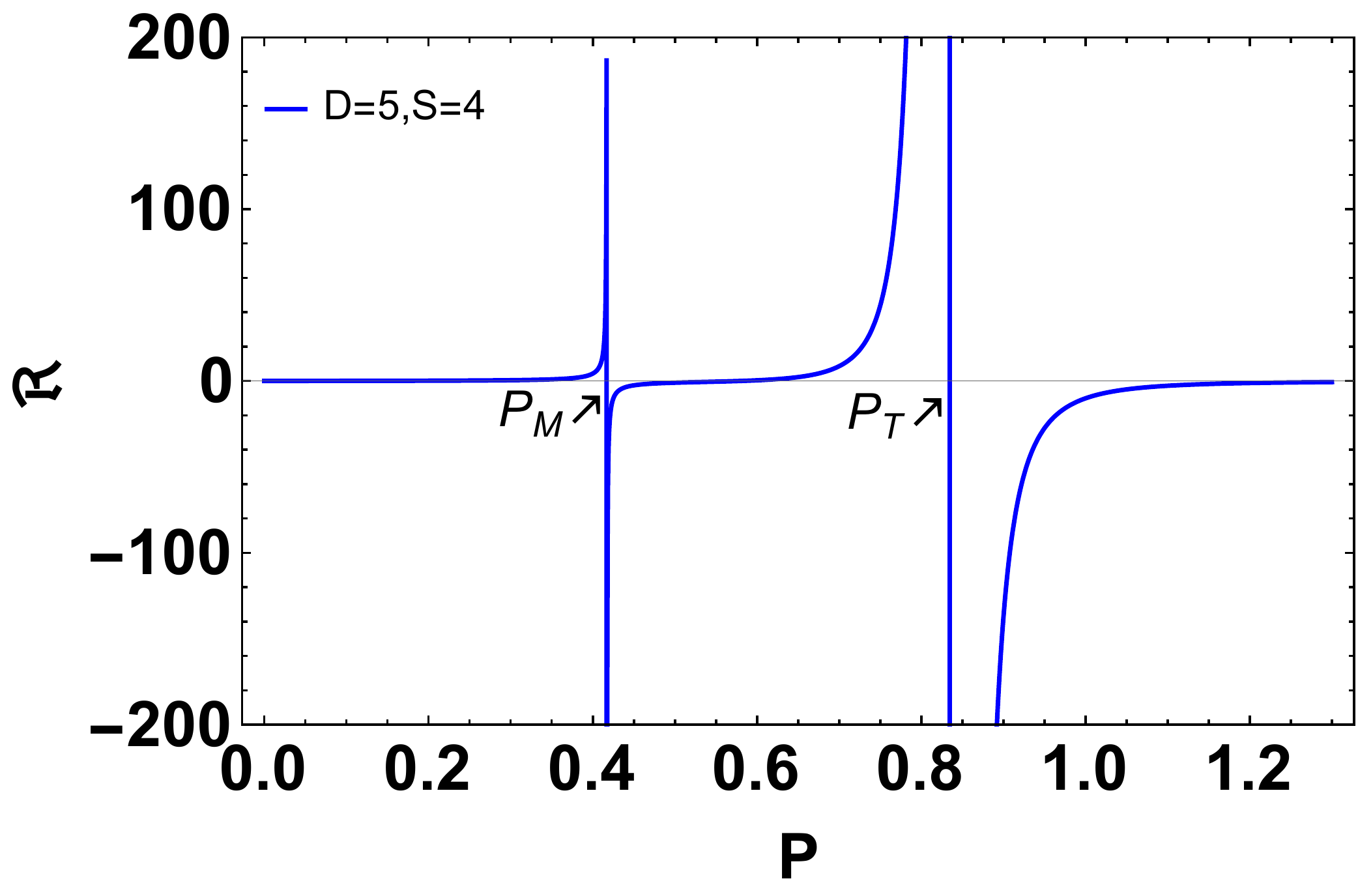}
		\caption{}
		\label{fig:2-2}
	\end{subfigure}
	\caption{The thermodynamic curvature of hyperbolic $\mathrm{AdS}$ black holes as a function of pressure for fixed value of entropy ($S=4$) in  $D=4$ (\ref{fig:2-1}) and  $D=5$ (\ref{fig:2-2}) dimensions.}
	\label{fig:2}
\end{figure}
It is obvious from Fig. (\ref{fig:2}) that the thermodynamic curvature is singular at two special pressure; $P_T$ and $P_M$, which are defined by Eqs. (\ref{pt}) and (\ref{pm}). Also, thermodynamic curvature as a function of temperature has been plotted in  Fig. (\ref{fig:3}). Thermodynamic curvature is singular at $T=0$ and $T=T_{min}$. We argued that $T=0$ corresponds to the $M=M_{min}$ because we show that
\begin{eqnarray}\label{TMMin}
T=\frac{(D-1)}{(D-2) S} \, (M-M_{min}).
\end{eqnarray}
Also, $T=T_{min}$ corresponds to $M=0$ which has been depicted in Fig. (\ref{fig:1}). In fact, the physical range with positive mass is $T\geq T_{min}$.
\begin{figure}[h]
	\begin{subfigure}{0.45\textwidth}\includegraphics[width=\textwidth]{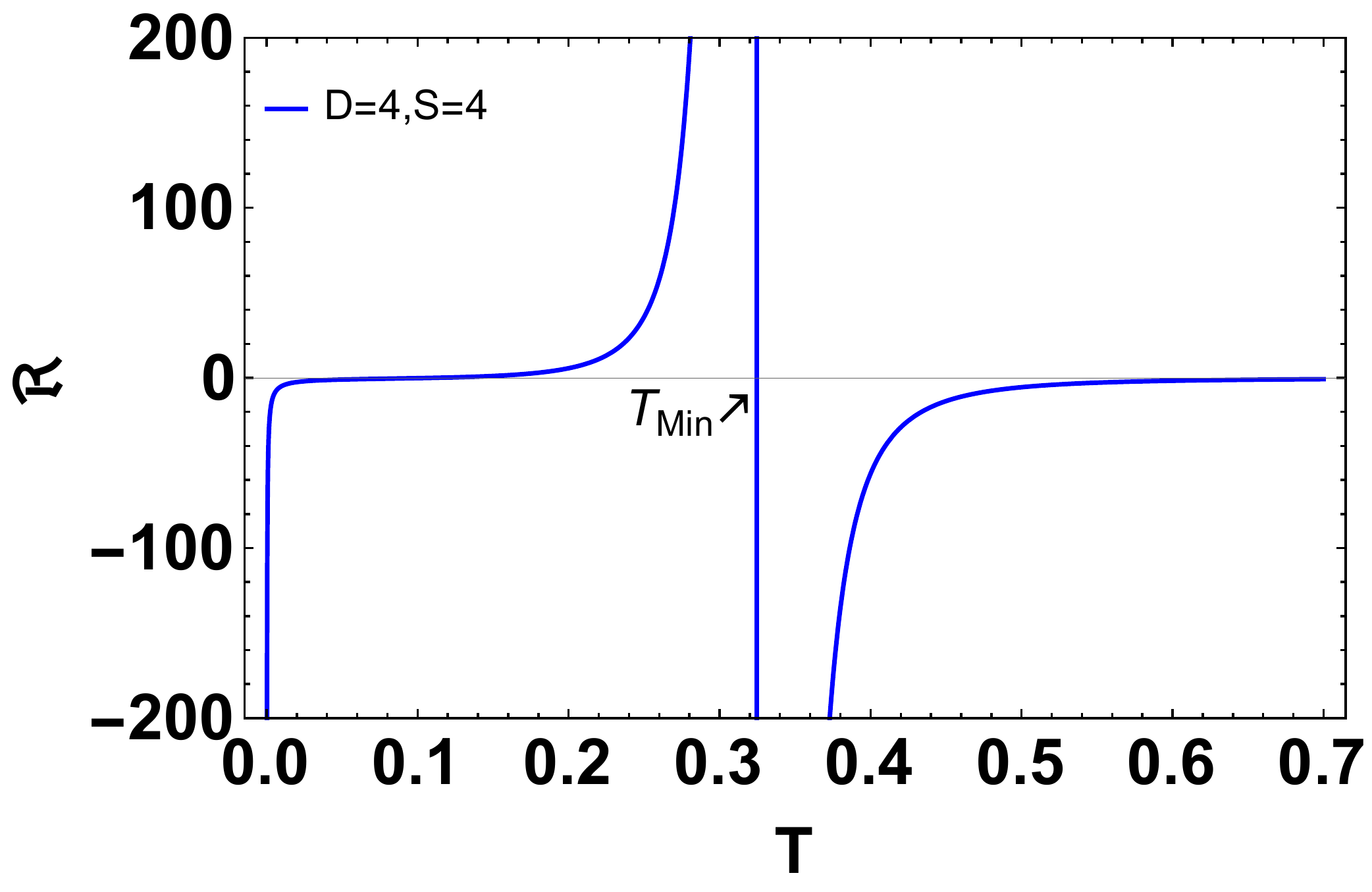}
		\caption{}
		\label{fig:3-1}
	\end{subfigure}
	\begin{subfigure}{0.45\textwidth}\includegraphics[width=\textwidth]{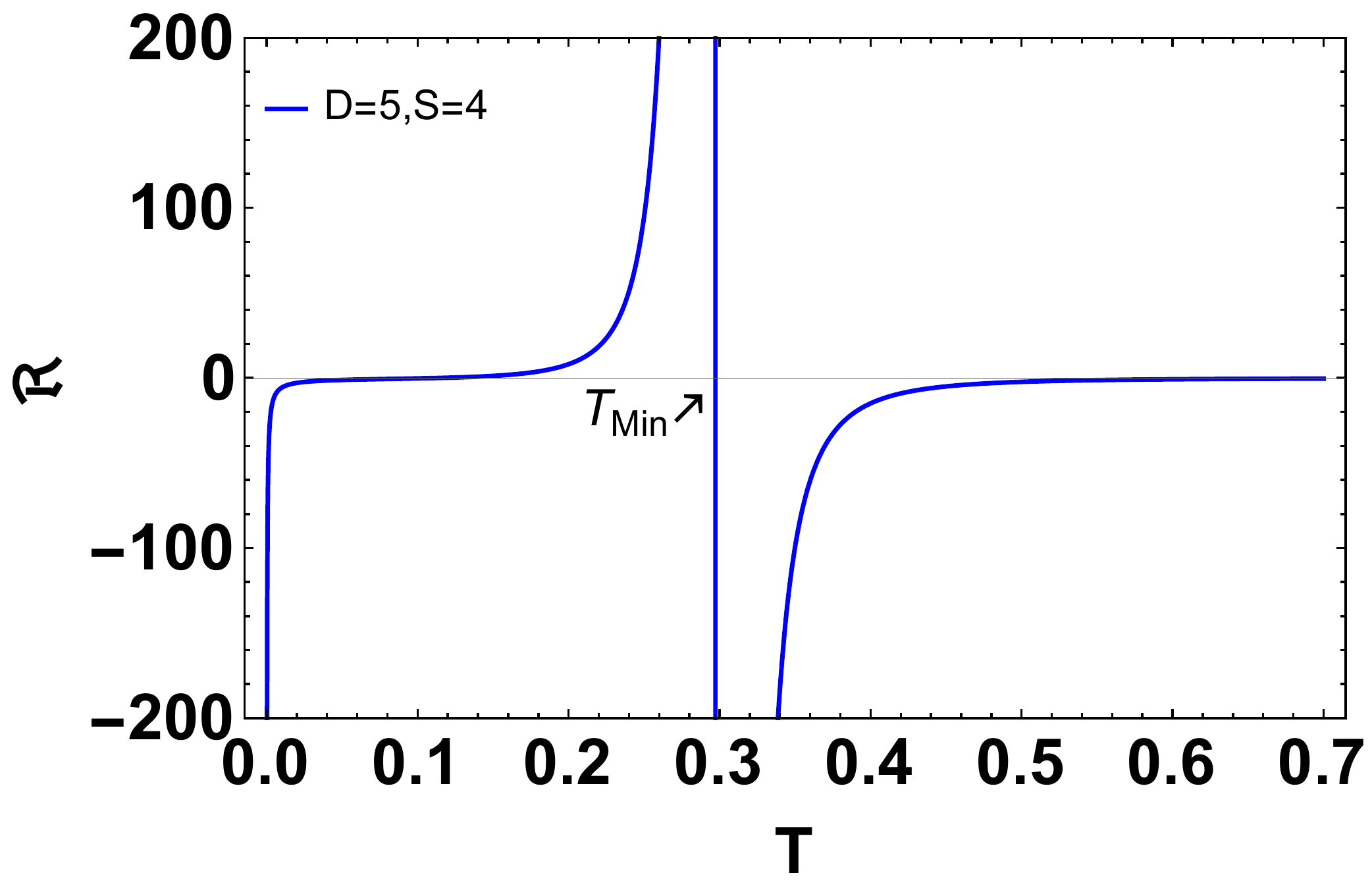}
		\caption{}
		\label{fig:3-2}
	\end{subfigure}
	\caption{Thermodynamic curvature as a function of temperature for fixed value of entropy $S=4$ for $D=4$ (\ref{fig:3-1}) and $D=5$ dimensions.}
	\label{fig:3}
\end{figure}
We focus on the singularity of thermodynamic curvature at $T_{min}$. By numerical methods, we show that the thermodynamic curvature in the vicinity of $T_{min}$ is singular as a power-law function in all dimensions 

 \begin{eqnarray}\label{RTmin}
\mathcal{R}\propto-(T-T_{min})^{-3}\,.
\end{eqnarray}
\end{widetext}

\subsection{Complexity}
\begin{figure}[h]
	\begin{subfigure}{0.45\textwidth}\includegraphics[width=\textwidth]{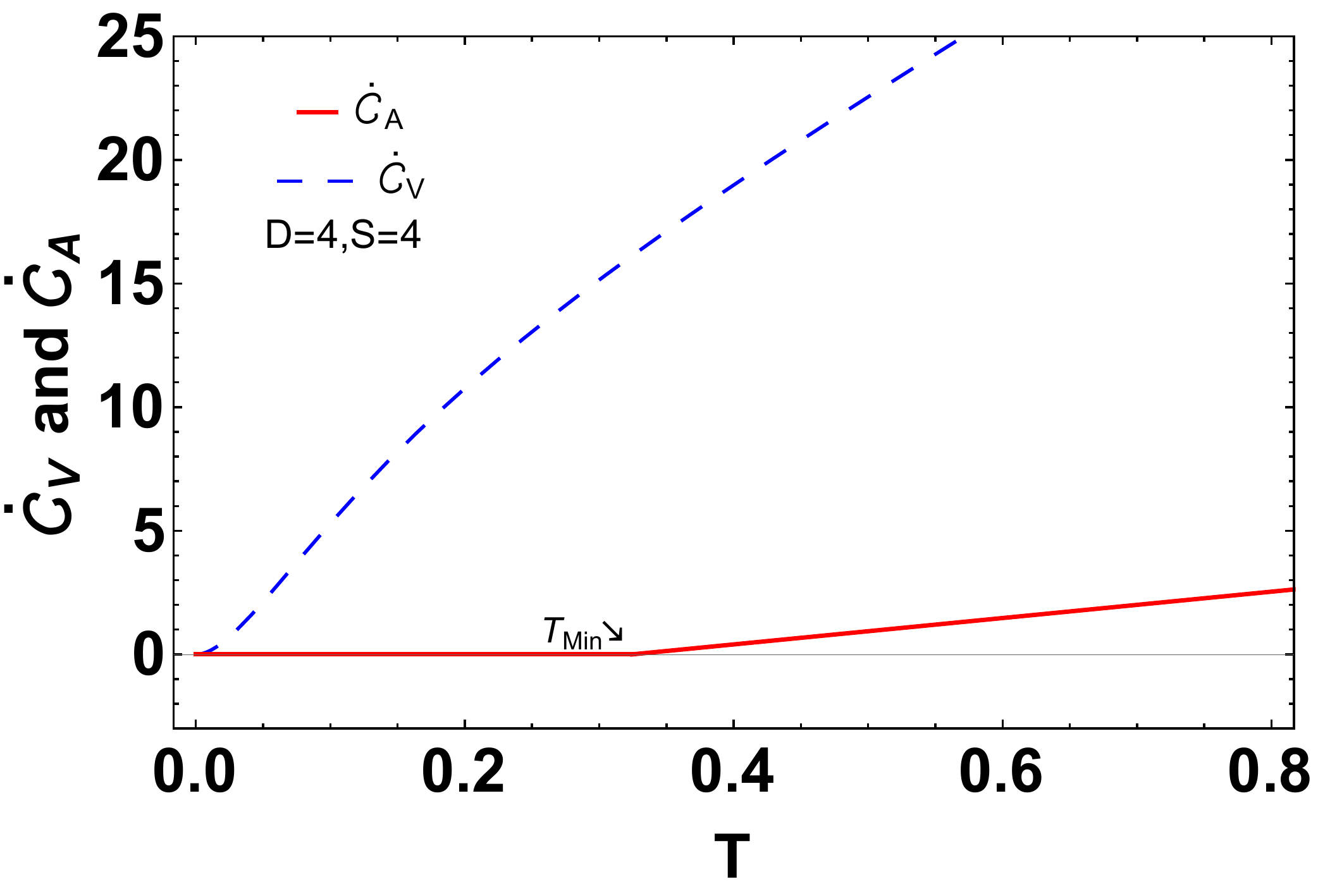}
		\caption{}
		\label{fig:4-1}
	\end{subfigure}
	\begin{subfigure}{0.45\textwidth}\includegraphics[width=\textwidth]{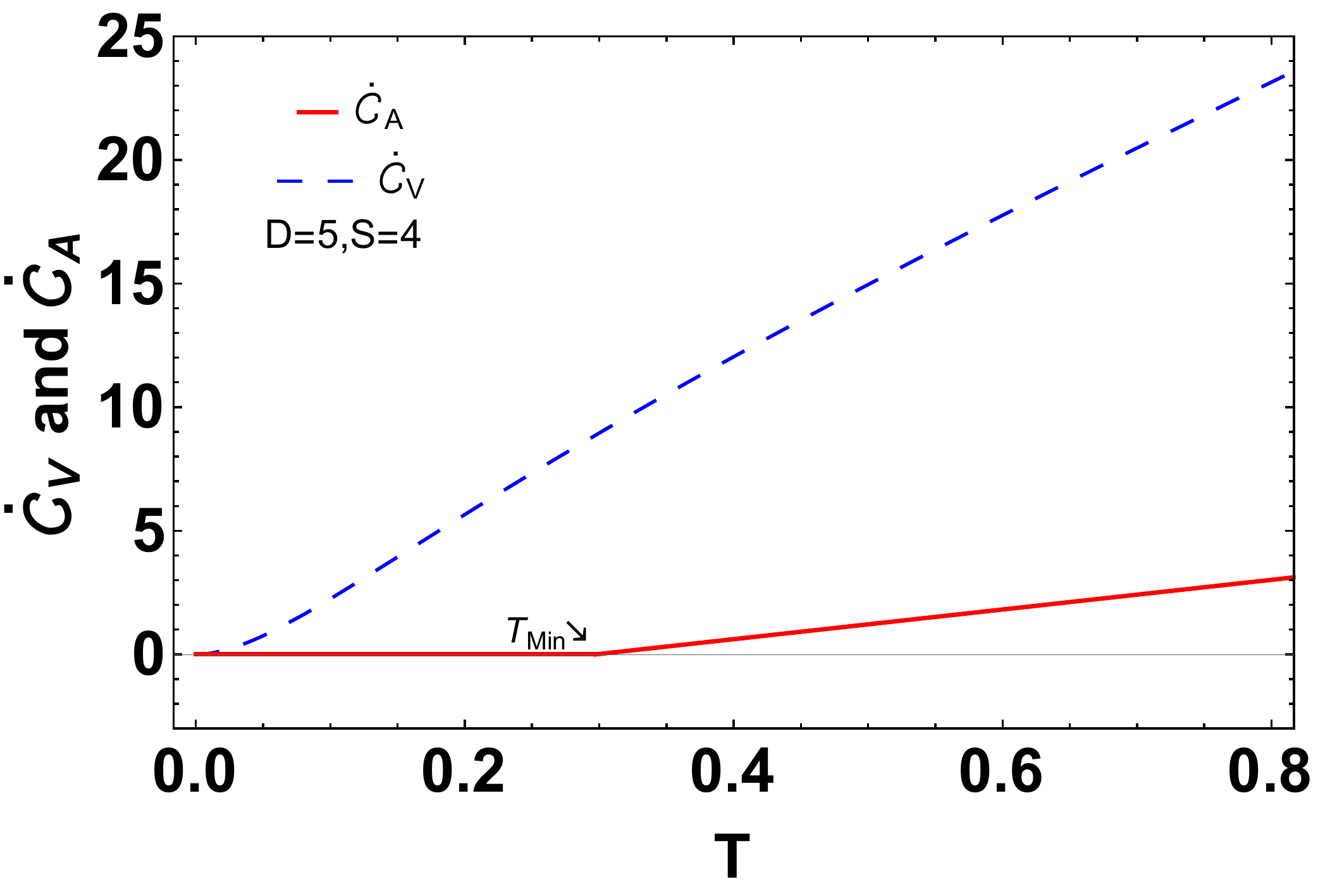}
		\caption{}
		\label{fig:4-2}
	\end{subfigure}
	\caption{The Complexity (equivalent to action; red solid line) and (equivalent to volume; blue dashed line) of as a function of temperature for fixed value of entropy $S=4$ in $D=4$ (\ref{fig:4-1}) and $D=5$ (\ref{fig:4-2}).}
	\label{fig:4}
\end{figure}

According to \cite{carmi2017time}, the complexity of hyperbolic black hole in $\mathrm{AdS}$ spacetime is obtained using two well-known conjectures. Using the action conjecture we obtain that
\begin{equation}\label{CAH}
\dot{\mathcal{C}}_A=2 M=\frac{2 (D-2)  S}{D-1}(T-T_{min}),
\end{equation}
and for $L\geq{r_h}$, the complexity vanishes $\dot{\mathcal{C}}_A=0$. The complexity as a function of temperature for a fixed value of entropy has been depicted in Fig. (\ref{fig:4}). It is obvious that the complexity based on the (CA) conjecture vanishes at $T=T_{min}$. We showed that the thermodynamic curvature is singular at $T=T_{min}$. In fact, one of the singular points of the thermodynamic curvature corresponds to zero complexity. 
The complexity is established 
\begin{equation}\label{CVH}
\dot{\mathcal{C}}_V \geq
\frac{8 \pi }{  D-2}(M-M_{min})\, ,
\end{equation}
which means that the complexity vanishes at some different point $M=M_{min}$. We showed that $M=M_{min}$ corresponds to $T=0$, where the thermodynamic curvature is singular.

Furthermore, we evaluate the heat capacity at constant pressure which is given by
\beq\label{Cp}
{{C}_{P}}=T{{\left( \frac{\partial S}{\partial T} \right)}_{P}}=\frac{T{{\left\{ S,P \right\}}_{S,P}}}{{{\left\{ T,P \right\}}_{S,P}}},
\eeq
where the last equality comes from the following relation
\beq
{{\left( \frac{\partial f}{\partial g} \right)}_{h}}=\frac{{{\left\{ f,h \right\}}_{a,b}}}{{{\left\{ g,h \right\}}_{a,b}}}\,,
\label{fgh}
\eeq
where the Nambu braket is defined \cite{mansoori2015hessian}
\beq
{{\{f,h\}}_{a,b}}={{\left( \frac{\partial f}{\partial a} \right)}_{b}}{{\left( \frac{\partial h}{\partial b} \right)}_{a}}-{{\left( \frac{\partial f}{\partial b} \right)}_{a}}{{\left( \frac{\partial h}{\partial a} \right)}_{b}}\label{fhab}.
\eeq
We obtain the heat capacity by evaluating the related Nambu brackets
\begin{eqnarray}\label{CpS}
C_{P}&=&\frac{( D-2) S \bigl(16 P \pi S^{\frac{2}{ D-2}} -  16^{\frac{1}{2 -  D}} ( D-3) ( D-2) \mathcal{V}^{\frac{2}{ D-2}}\bigr)}{16 P \pi S^{\frac{2}{ D-2}} + 16^{\frac{1}{2 -  D}} ( D-3) ( D-2) \mathcal{V}^{\frac{2}{D-2}}}\nonumber\\
&=&( D-2)\frac{ (P -  P_T) }{P+P_T} S\,.
\end{eqnarray}
The above equation shows that the heat capacity at constant pressure vanishes at $P=P_{T}$ it corresponds to $T=0$ where the thermodynamic curvature is singular. The heat capacity as a function of temperature obtained 
\begin{eqnarray}\label{CpT}
C_{P}&=&\frac{2 \pi  ( D-2) S^{1 + \frac{1}{ D-2}} T}{  4^{\frac{1}{2 -  D}} ( D-3) \mathcal{V}^{\frac{1}{ D-2}}+2 \pi S^{\frac{1}{ D-2}} T}.
\end{eqnarray}
It is obvious from Eq.\ref{CpT} that the heat capacity vanished at $T=0$.
\section{ black holes with momentum relaxation}\label{3}
It is well-known that the holographic models with broken translational symmetry yield theories with momentum relaxation. The Einstein-Maxwell action in $(d+1)$-dimensional spacetime is supplemented by $d-1$ massless 
scalar fields that break the translational invariance of the boundary theory in the context of the $\mathrm{AdS/CFT}$ duality. In fact, scalar axion fields 
enter the bulk action only through the kinetic term and the sources are linear in the boundary.  In the following, we will consider the thermodynamic, thermodynamic geometry, and the complexity of black holes with the momentum relaxation.
\subsection{Thermodynamic of black holes }
Though the Einstein-Maxwell-Axion gravity is a well-known theory \cite{andrade2014simple}, in this paper we focus on the Einstein-Axion theory with the total action
\begin{equation}
S=\frac{1}{16\pi }\int \!  \sqrt{-g} \,\bigg( R-2\Lambda -\frac{1}{2}\sum_{I=1}^{D-2}(\partial\psi_I)^2 \bigg) \,d^Dx ,
\label{eq:actionBeta}
\end{equation}
where $D$ is the dimension of spacetime, $\psi_{I}$ is the axion field $D-2$ dimensions via the last kinetic term in the action. One can suppose that the scalar fields linearly depend on the $D-2$ dimensional spatial coordinates $x^{a}$, i.e., $\psi_{i}=\beta\delta_{Ia}x^{a}$. The action admits the following black hole solution
\begin{eqnarray}\label{eq-metricBeta}
ds^2=-f(r)dt^2+\frac{1}{f(r)}dr^2+r^2dx^a dx^a,\end{eqnarray}
where 
\begin{eqnarray}\label{fr}
f(r)=\frac{r^2}{L^2}-\frac{\beta^2}{2(D-3)}-\frac{m_0}{r^{D-3}},
\end{eqnarray}
and 
\begin{eqnarray}\label{eq-mQBeta}
M&=&\frac{(D-2)\mathcal{V}_{D-2}}{16\pi}m_0 \,\,,\\
m_0&=&\frac{16 \pi  P r_h^{D-1}}{(D-1) (D-2)}-\frac{\beta^2 r_h^{D-3}}{2(D-3)},\nonumber
\end{eqnarray}
where $m_0$ is a mass parameter and $\mathcal{V}_{D-2}$ is the dimensionless volume of $D-2$-dimensional spatial geometry $x^a$. We can evaluate the temperature 
\begin{eqnarray}\label{eq-TBeta}
T&=&\frac{f'(r_h)}{4\pi} \nonumber\\
&=&\frac{1}{4\pi}\left(\frac{(D-1) r_h}{L^2}-\frac{\beta ^2}{2r_h}\right).
\end{eqnarray}
The entropy and the pressure of the black hole are given by
\begin{equation}\label{SPeq}
S=\frac{\mathcal{V}_{D-2}}{4}r_h^{D-2}\,\,\,, \,\,\,P=-\frac{\Lambda}{8\pi}=\frac{(D-1)(D-2)}{16\pi L^2}.
\end{equation}
We inspect that the above physical quantities satisfy the following first law of black hole thermodynamic. The first thermodynamic law is given by
\begin{equation}\label{eq-dMBeta}
dM=T dS+V dP+\varphi d\beta,
\end{equation}
where $\varphi$ is the conjugate variable to the strength of momentum relaxation with definition  
\begin{equation}\label{eq-phiBeta}
\varphi=\frac{\partial M}{\partial \beta}\big|_{S,P}=-\frac{(D-2)\beta\mathcal{V}_{D-2}r_h^{D-3}}{16\pi(D-3)},
\end{equation}
and the Smarr relation for the black hole is obtained 
\begin{equation}\label{eq-SmarrBeta}
(D-3)M=(D-2)TS-2PV.
\end{equation}
\subsection{Thermodynamic geometry}
We study the thermodynamic geometry of a black hole with momentum relaxation. First, we do not consider the momentum relaxation parameter, $\beta$, as a fluctuating thermodynamic parameter. It means that the momentum relaxation is considered as a constant and we will consider a two-dimensional thermodynamic parameters space with fluctuating parameters entropy ($S$) and pressure ($P$). It is straightforward to obtain the metric tensor components of thermodynamic parameters space. Using the evaluated metric elements, we work out the thermodynamic curvature 
\begin{widetext}
	\begin{eqnarray}\label{eq-RBeta}
	\mathcal{R}= \frac{a_0 \pi S^{-2 + \frac{1}{D-2}} \beta^6+a_1 \pi^2 P S^{-2 + \frac{3}{ D-2}} \beta^4-a_2 3\pi^3  P^2 S^{2 + \frac{5}{ D-2}} \beta^2}{\biggl(- 2^{5 + \frac{4}{ D-2}} P \pi S^{\frac{2}{D-2}} + ( D-2) \beta^2\biggr) \biggl(- 2^{5 + \frac{4}{ D-2}} ( D-3) P \pi S^{\frac{2}{D-2}} + ( D-2) ( D-1) \beta^2\biggr)^3},
	\end{eqnarray}
where
		\begin{subequations}
				\begin{align} 
			a_0=2^{4 + \frac{2}{ D-2}} ( D-3) ( D-2)^2 ( D^2-1)\,,     
			\end{align}
		\begin{align} 
         	a_1=2^{10 + \frac{6}{ D-2}} ( D-3)^2 ( D-2) ( D-1)\,,    
		\end{align}
		\begin{align} 
		a_2= 2^{14 + \frac{10}{ D-2}} ( D-3)^2 (D-1)\,,    
		\end{align}
	\end{subequations}
where we set $\mathcal{V}_{D-2}=1$. Using the definition of total entropy, total mass and temperature of the black hole by Eqs. (\ref{eq-mQBeta}), (\ref{eq-TBeta}) and (\ref{SPeq}), we plotted $M$ and $T$	versus $\beta$ for fixed values of entropy and pressure in $D=4$ and $D=5$ in Fig. (\ref{fig:6}).
\begin{figure}[h]
	\begin{subfigure}{0.45\textwidth}\includegraphics[width=\textwidth]{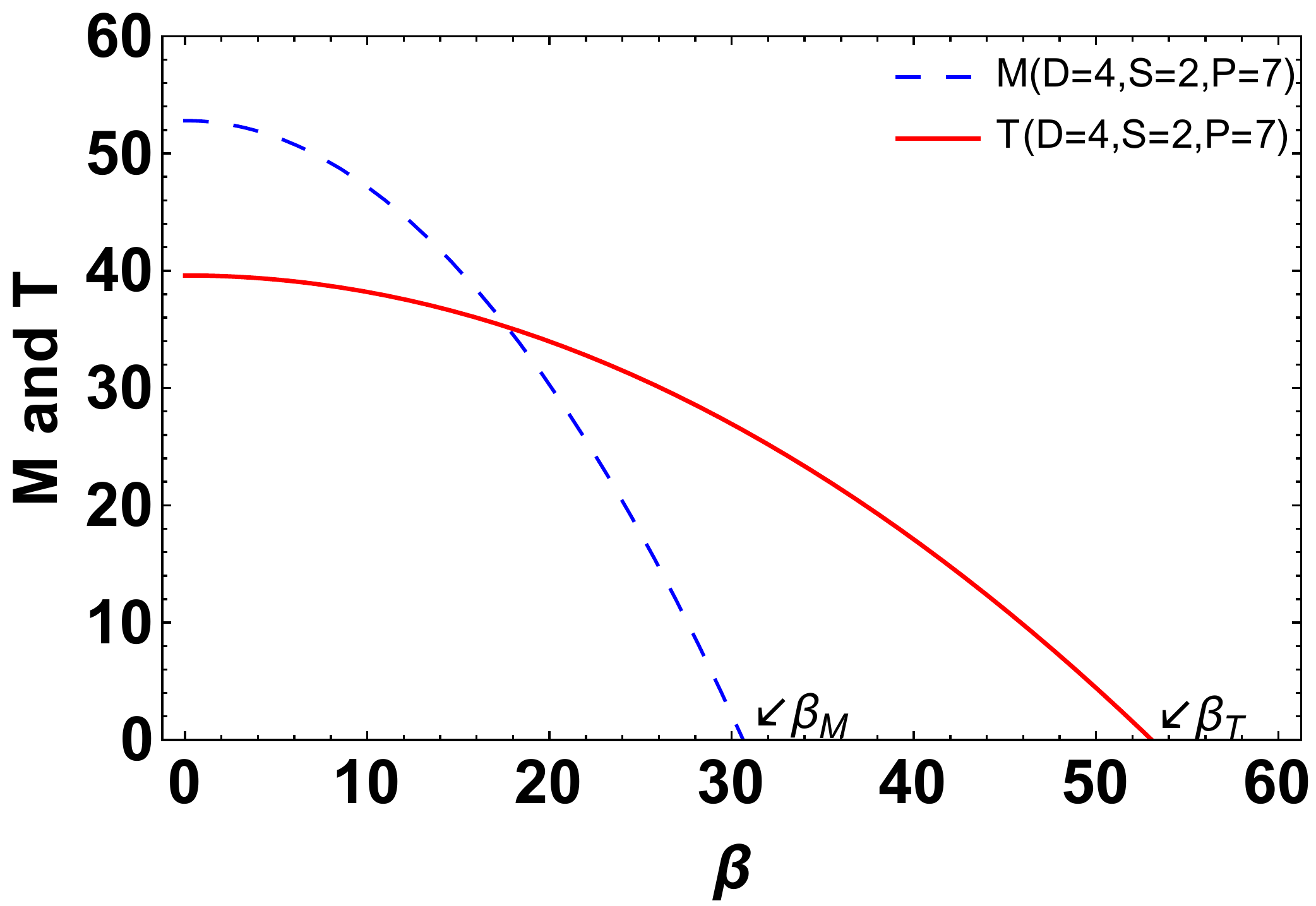}
		\caption{}
		\label{fig:6-1}
	\end{subfigure}
	\begin{subfigure}{0.45\textwidth}\includegraphics[width=\textwidth]{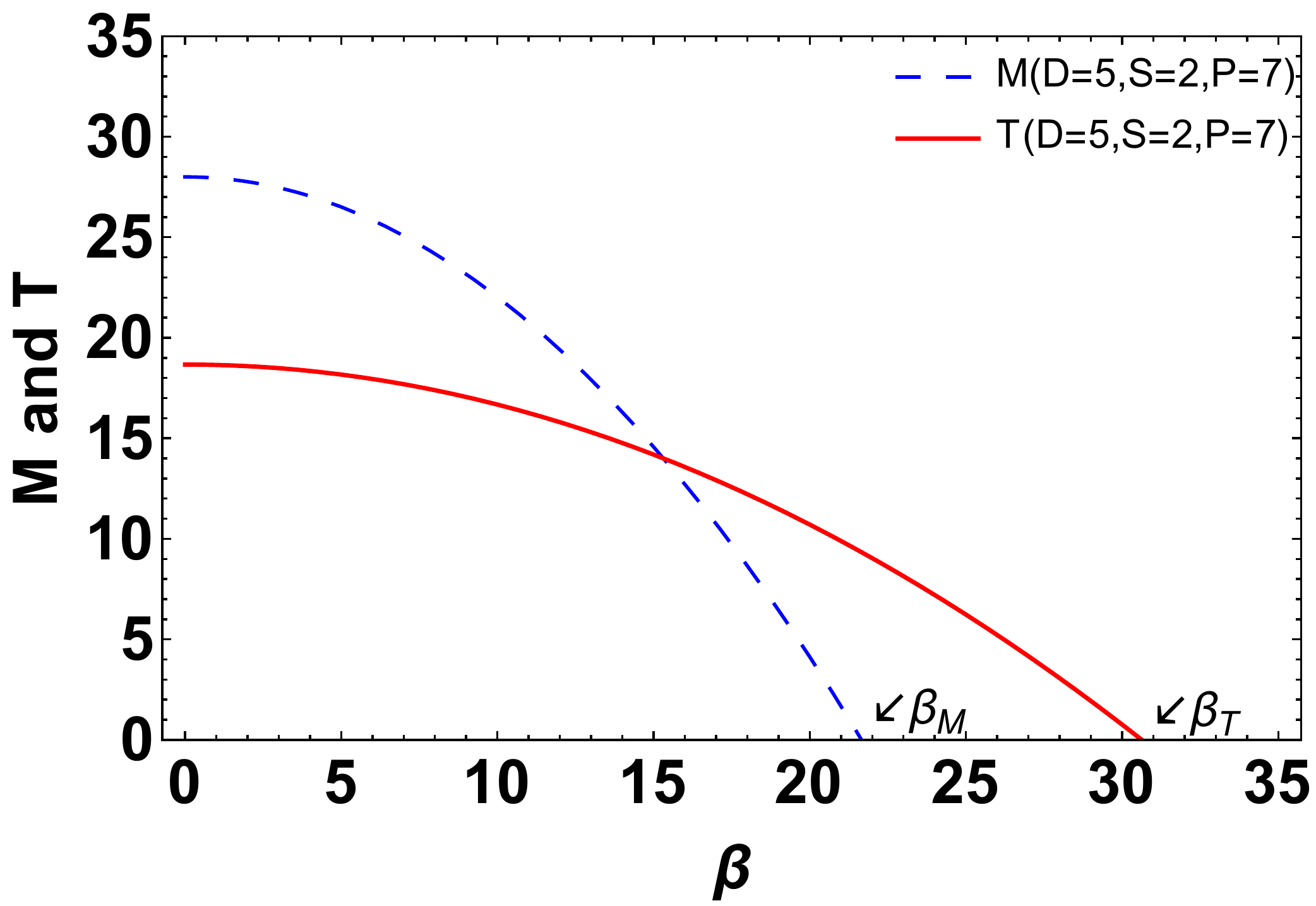}
		\caption{}
		\label{fig:6-2}
	\end{subfigure}
	\caption{Mass  (blue, dashed line) and temperature (red, solid line) of black hole as a function of momentum relaxation parameter for fixed values of entropy ($S=4$) and pressure ($P=7$) in $D=4$ (\ref{fig:6-1}) and $D=5$ (\ref{fig:6-2}).}
	\label{fig:6}
\end{figure}
We can obtain two special values $\beta_{T}$ and $\beta_{M}$ for momentum relaxation parameter which correspond to $T=0$ and $M=0$ respectively which are given by

\begin{subequations}
	\begin{align}\label{betaT}
	\beta^2_{T}=\pi \frac{2^{5 + \frac{4}{ D-2}}}{ D-2} P  S^{\frac{2}{ D-2}},     
	\end{align}
	\begin{align} \label{betaM}
	\beta^2_{M}=\pi \frac{2^{5 + \frac{4}{ D-2}} ( D-3)}{( D-2) ( D-1)} P  S^{\frac{2}{ D-2}}.  
	\end{align}
\end{subequations}
These special values of $\beta$ are clear in Fig. (\ref{fig:6}). In fact, for $\beta>\beta_{T}$, the mass and temperature will be negative. For $\beta_{M}<\beta<\beta_{T}$, the temperature is positive while mass is negative, and finally for $\beta<\beta_{M}$, both of the temperature and mass are positive.

The thermodynamic curvature has been depicted in Fig. (\ref{fig:5}) as a function of momentum relaxation parameter for fixed values of the entropy and pressure in $D=4$ and $D=5$ dimensions. It is obvious that the thermodynamic curvature is singular at the special values of $\beta=\beta_{T}$ and $\beta=\beta_{M}$.
\begin{figure}[h]
	\begin{subfigure}{0.45\textwidth}\includegraphics[width=\textwidth]{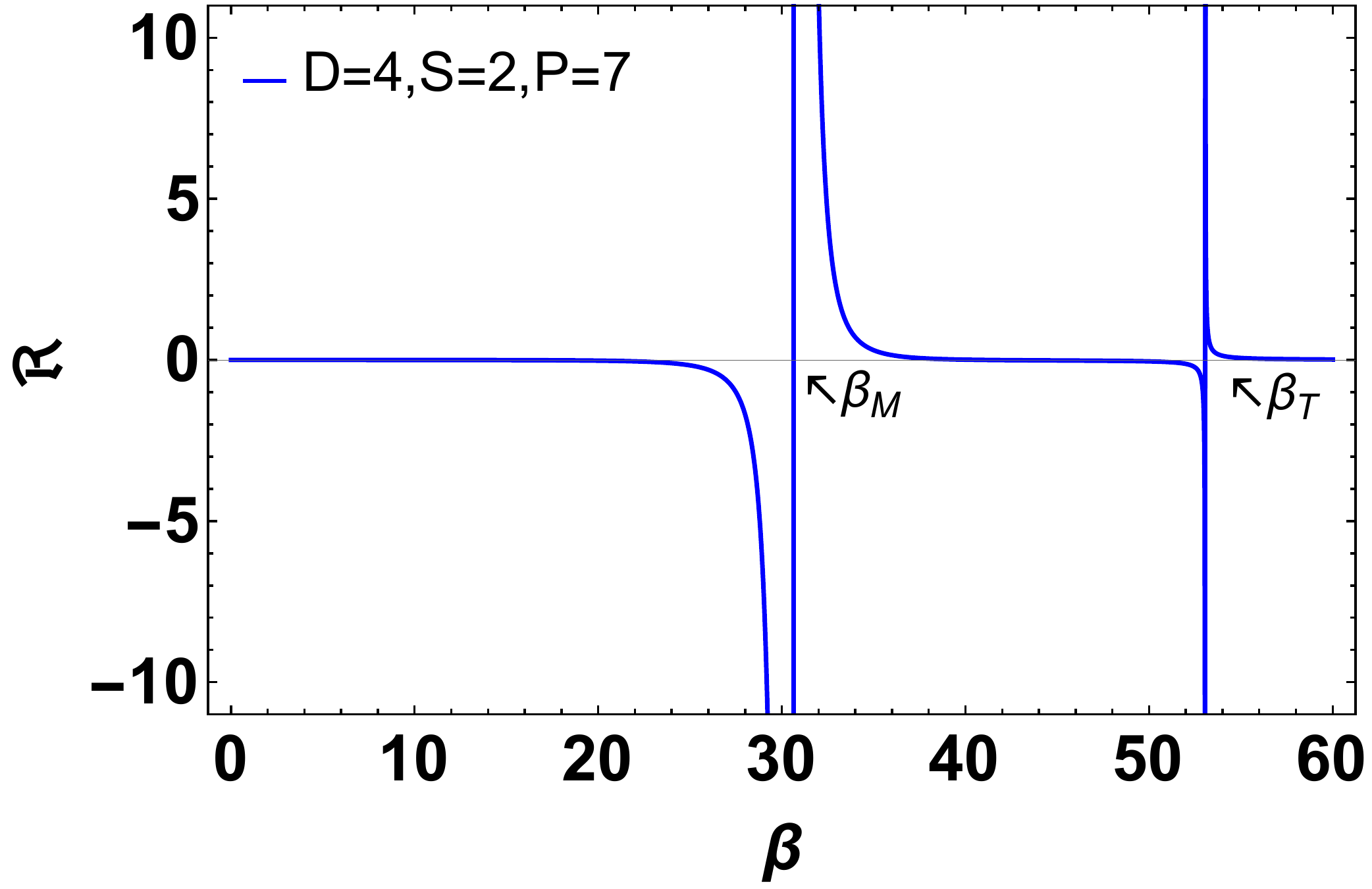}
		\caption{}
		\label{fig:5-1}
	\end{subfigure}
	\begin{subfigure}{0.45\textwidth}\includegraphics[width=\textwidth]{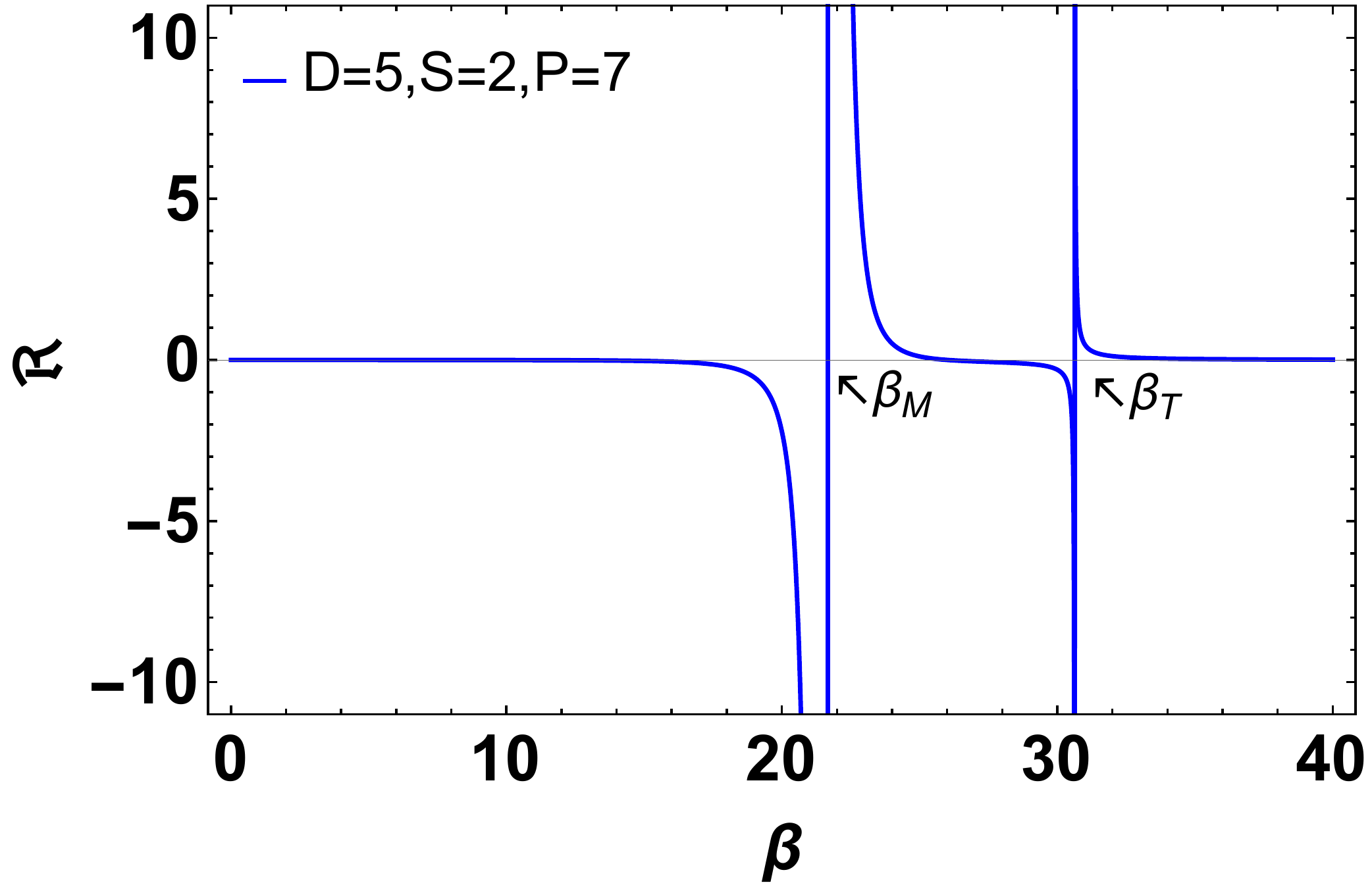}
		\caption{}
		\label{fig:5-2}
	\end{subfigure}
	\caption{Thermodynamic curvature of two dimensional thermodynamic parameters space for a black hole with momentum relaxation as a function of momentum relaxation parameter for fixed values of entropy ($S=4$) and pressure ($P=7$) in $D=4$ (\ref{fig:5-1}) and $D=5$ (\ref{fig:5-2}).}
	\label{fig:5}
\end{figure}
We could extract the thermodynamic curvature as a function of temperature. First, we obtain $\beta$ as a function of $S$, $P$ and $T$ using Eqs. (\ref{eq-TBeta}) and (\ref{SPeq})
\begin{equation}\label{BetaSPT}
\beta^2=2^{3 + \frac{2}{ D-2}} \pi S^{\frac{1}{ D-2}}\bigg(\frac{2^{2 + \frac{2}{ D-2}} P S^{\frac{1}{D-2}}}{ D-2} -T\bigg)\,.
\end{equation}
We mentioned that $\beta_{M}$ corresponds to $M=0$. Using Eqs. (\ref{betaM}) and (\ref{BetaSPT}), we obtain a minimum value for temperature as follows:
\begin{equation}\label{TminR3beta}
T_{min}= \frac{2^{3 + \frac{2}{D-2}} }{(D-2) (D-1)}P S^{\frac{1}{D-2}}\,.
\end{equation}
In fact, for $T<T_{min}$, the black hole mass will be negative and $T=T_{min}$ corresponds to $M=0$.
According to the above equations, rewrite the thermodynamic curvature (\ref{eq-RBeta}) as a function of the temperature 
	\begin{equation}\label{RBetaTPS}
	\mathcal{R}= \frac{ \bigl( b_0 P^3 S^{\frac{3}{D-2}} -b_1 P^2 S^{\frac{2}{D-2}} T +b_2 P S^{\frac{1}{ D-2}} T^2 -b_3 T^3\bigr)}{( D-2)^5 ( D-1)^2 S^2 T\,\, (T-T_{min})^3},~~~~~~
	\end{equation}
where $b_{0}$, $b_{1}$, $b_{2}$, and $b_{3}$ are dimension dependent coefficients and given by

\begin{subequations}
	\begin{align}\label{a0Beta}
	b_0=2^{9 + \frac{6}{ D-2}} ( D-3),     
	\end{align}
	\begin{align} 
	b_1=2^{7 + \frac{4}{ D-2}} ( D-3) ( D-2) D,  
	\end{align}
		\begin{align} 
	b_2=2^{3 + \frac{2}{ D-2}} ( D-3) ( D-2)^2 ( 5 D-3),  
	\end{align}
		\begin{align} 
	b_3=2 (D-2)^3 ( D+1) ( D-3). 
	\end{align}
\end{subequations}
Thermodynamic curvature as a function of temperature has been plotted in Fig. (\ref{fig:7}). This figure indicates that the thermodynamic curvature is singular at $T=T_{min}$ and $T=0$. For the physical range $T>T_{min}$, mass is positive and the thermodynamic curvature is negative. 
\begin{figure}[h]
	\begin{subfigure}{0.45\textwidth}\includegraphics[width=\textwidth]{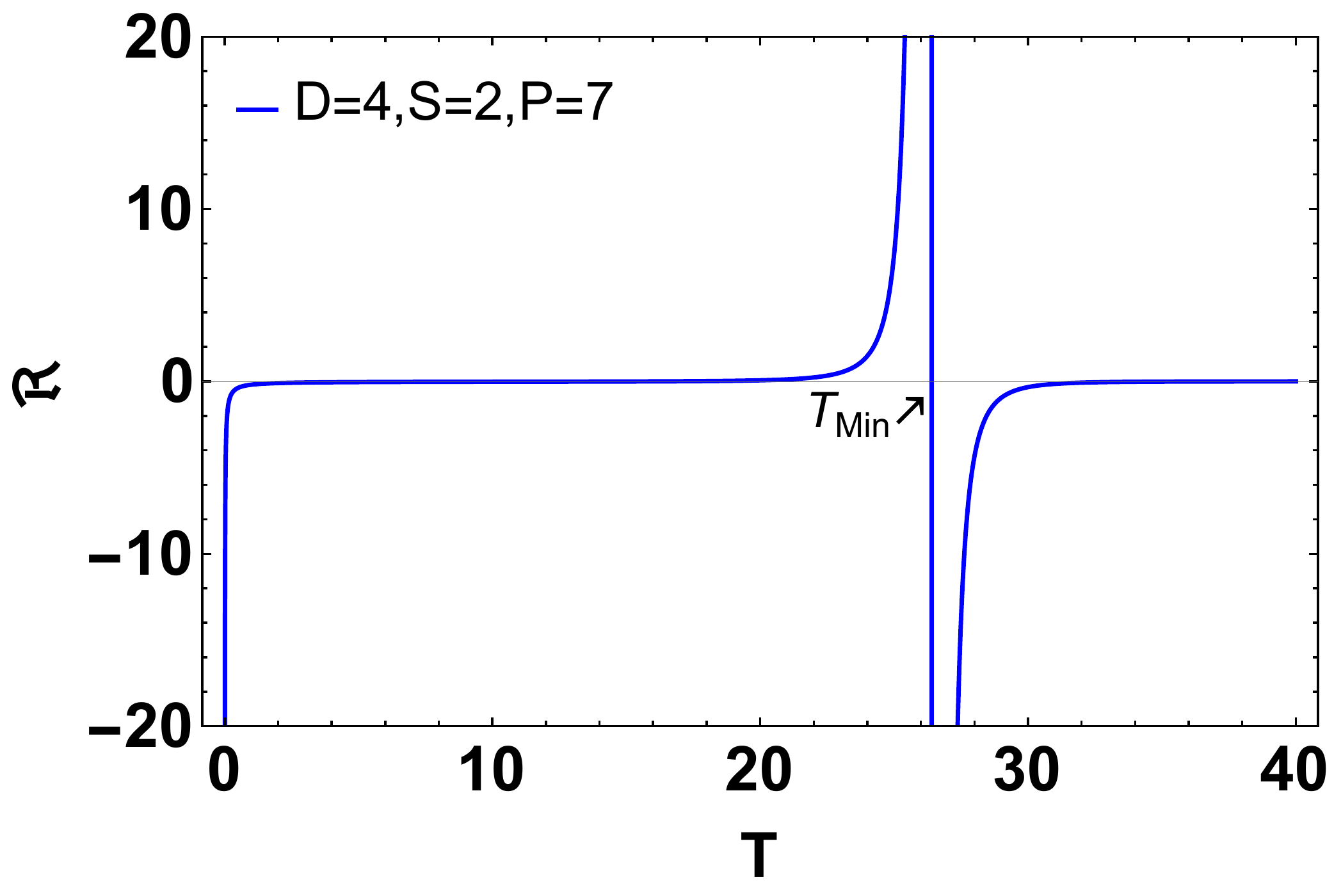}
		\caption{}
		\label{fig:7-1}
	\end{subfigure}
	\begin{subfigure}{0.45\textwidth}\includegraphics[width=\textwidth]{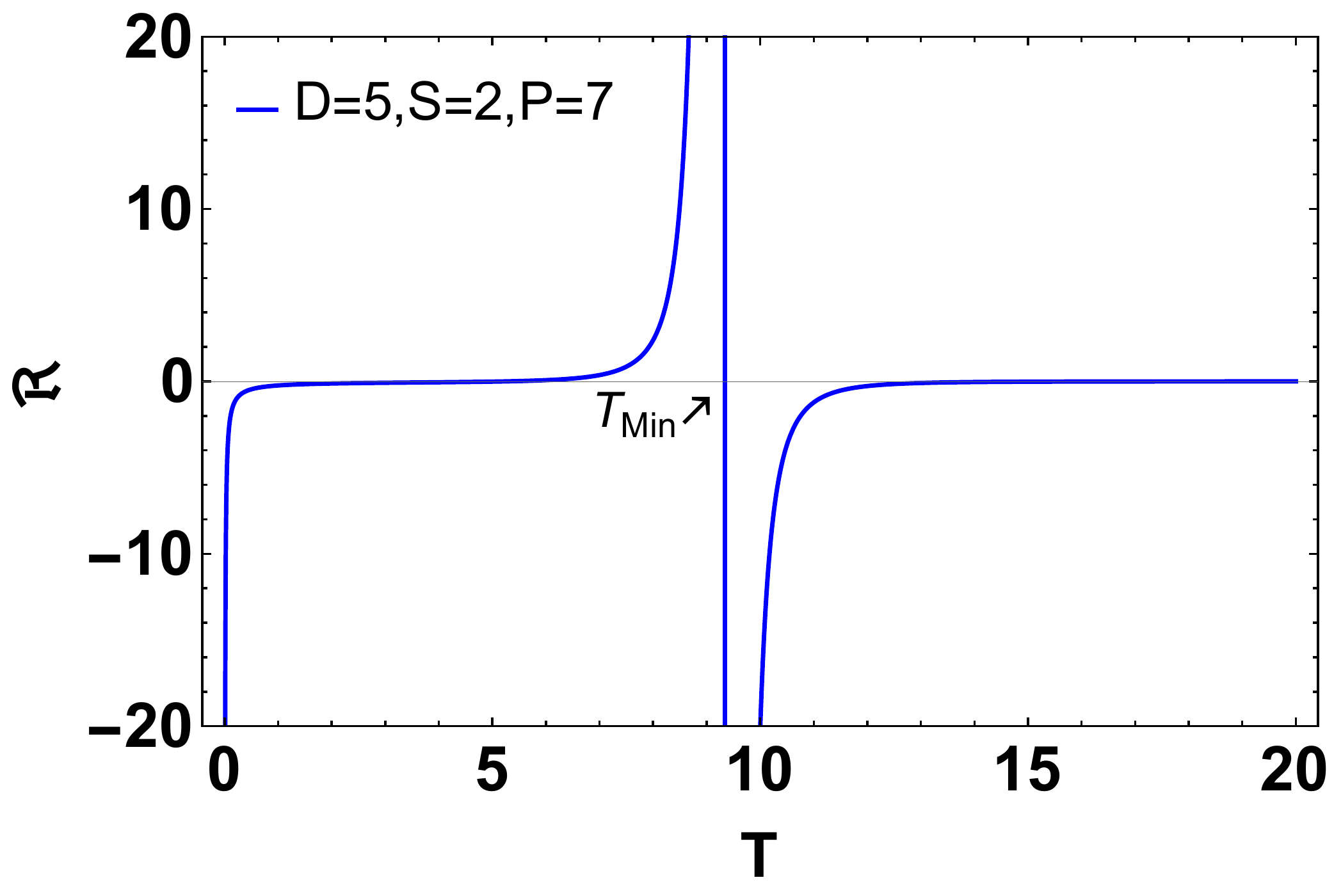}
		\caption{}
		\label{fig:7-2}
	\end{subfigure}
	\caption{Thermodynamic curvature of black hole with momentum relaxation as a function of temperature for fixed values of entropy ($S=2$) and Pressure ($P=7$) in $D=4$ (\ref{fig:7-1}) and $D=5$ (\ref{fig:7-2})}
	\label{fig:7}
\end{figure}
Also, we obtain that the thermodynamic curvature has a power-law behaviour in the vicinity of $T=T_{min}$ as follows:
	\begin{equation}\label{RbetaTmin}
\mathcal{R}\propto -(T-T_{min})^{-3}\,.
\end{equation}
\subsection{Thermodynamic geometry  with fluctuations of   momentum relaxation}
It has been shown that in $\mathrm{AdS}$ black holes with axionic charge \cite{caldarelli2017phases,fang2017holographic}, It is necessary to include some parameters to the set of fluctuating thermodynamic parameters to obtain a relevant first thermodynamic law. Consideration of the thermodynamic geometry in extended phase space, contains some useful information and interesting results. We considered the thermodynamic geometry in a two dimensional thermodynamic parameters space. By including the momentum relaxation parameter to the set of thermodynamic fluctuating parameters, we will consider the thermodynamic curvature of the extended thermodynamic phase space. We obtain the components of three dimensional thermodynamic parameters space using Eq. (\ref{Ruppeiner}). We notice that the thermodynamic parameters are $X^1=S$, $X^2=P$ and $X^3=\beta$. Evaluation of the thermodynamic curvature using the nine components of metric tensor of three dimensional thermodynamic parameters space is straightforward. We obtain that
	\begin{equation}\label{R3SPbeta}
	\mathcal{R}= \frac{c_0P^3 \pi^4 S^{ \frac{7}{ D-2}-2}+c_1P^2 \pi^3 S^{\frac{5}{ D-2}-2} \beta^2 +c_2P \pi^2 S^{ \frac{3}{ D-2}-2} \beta^4+c_3\pi S^{ \frac{1}{D-2}-2} \beta^6}{( D-2) \bigl(2^{5 + \frac{4}{ D-2}} P \pi S^{\frac{2}{ D-2}} -  ( D-2) \beta^2\bigr) \bigl(2^{5 + \frac{4}{ D-2}} ( D-3) P \pi S^{\frac{2}{ D-2}} -  ( D-2) (D-1) \beta^2\bigr)^3}\,,
	\end{equation}
where the coefficients $c_0$, $c_1$, $c_2$ and $c_3$ are given by
\begin{subequations}
	\begin{align}\label{a0R3}
c_0=2^{20 + \frac{12}{ D-2}} ( D-3) ( D-1) \bigl( 4^{\frac{1}{ D-2}} (9 + D^2)-3 D \times 2^{\frac{D}{ D-2}} \bigr) \, ,   
	\end{align}
	\begin{align} 
c_1=- 2^{14 + \frac{10}{D-2}} (D-3)^2 ( D-2) ( D-1) ( 4 D-9)\, ,  
	\end{align}
	\begin{align} 
c_2=2^{-\frac{26}{ D-2}} ( D-3) ( D-1) \Biggl( 2^{9 + \frac{32}{ D-2}} \biggl(148 + D \Bigl( D \bigl(245 + 2 ( D-13) D\bigr)-220 \Bigr)\biggr)- 2^{\frac{16 D}{ D-2}} D^2 \Biggr)\, ,  
	\end{align}
	\begin{align} 
	c_3=&2^{4 - \frac{10}{ D-2}} (D-3) ( D-1) \Biggl( 4^{\frac{3 D}{ D-2}} \bigl( D (7 + D + 3 D^2)-3 \bigr) + 4^{\frac{6}{ D-2}} \biggl(336 + D \Bigl( D \bigl(152 + D ( 53 D-264 )\bigr)-736 \Bigr)\biggr)\nonumber\\
&-11\times 4^{\frac{4 + D}{ D-2}} D^4 \Biggr)\,.
	\end{align}
\end{subequations}
The thermodynamic curvatures of extended thermodynamic phase space for $D=4$ and $D=5$ dimensional solutions are different. For $D=4$ dimensions, the thermodynamic curvature is given by
\begin{equation}\label{R3D4}
\mathcal{R}_{D=4}=\frac{48 \pi (32 \pi P S -  \beta^2) (128 \pi P S + 3  \beta^2)}{S^{3/2} (64 \pi P S - 3  \beta^2)^2 (64 \pi P S -  \beta^2)}\,,
\end{equation}
and for $D=5$ dimensions, we obtain that
\begin{equation}\label{R3D5}
\mathcal{R}_{D=5}=\frac{2 \pi (4194304 2^{2/3} P^3 \pi^3 S^2 - 540672 2^{1/3} P^2 \pi^2 S^{4/3} \beta^2 - 3456 P \pi S^{2/3} \beta^4 + 729 2^{2/3} \beta^6)}{3 S^{5/3} (32 2^{1/3} P \pi S^{2/3} - 3 \beta^2)^3 (64 2^{1/3} P \pi S^{2/3} - 3 \beta^2)}\,,
 \end{equation}
It is obvious that the dependence of thermodynamic curvature to thermodynamic parameters is different in $D=4$ and $D=5$. More especially, the appearance of the term $\beta^{6}$  in the numerator is related to the five and higher dimensional solutions. 

Thermodynamic geometry as a function of momentum relaxation parameter for fixed values of entropy and pressure have been plotted in (\ref{fig:9}). It can be observed that the thermodynamic curvature is singular at $\beta_{M}$ and $\beta_T$. 
 \begin{figure}[h]
 	\begin{subfigure}{0.45\textwidth}\includegraphics[width=\textwidth]{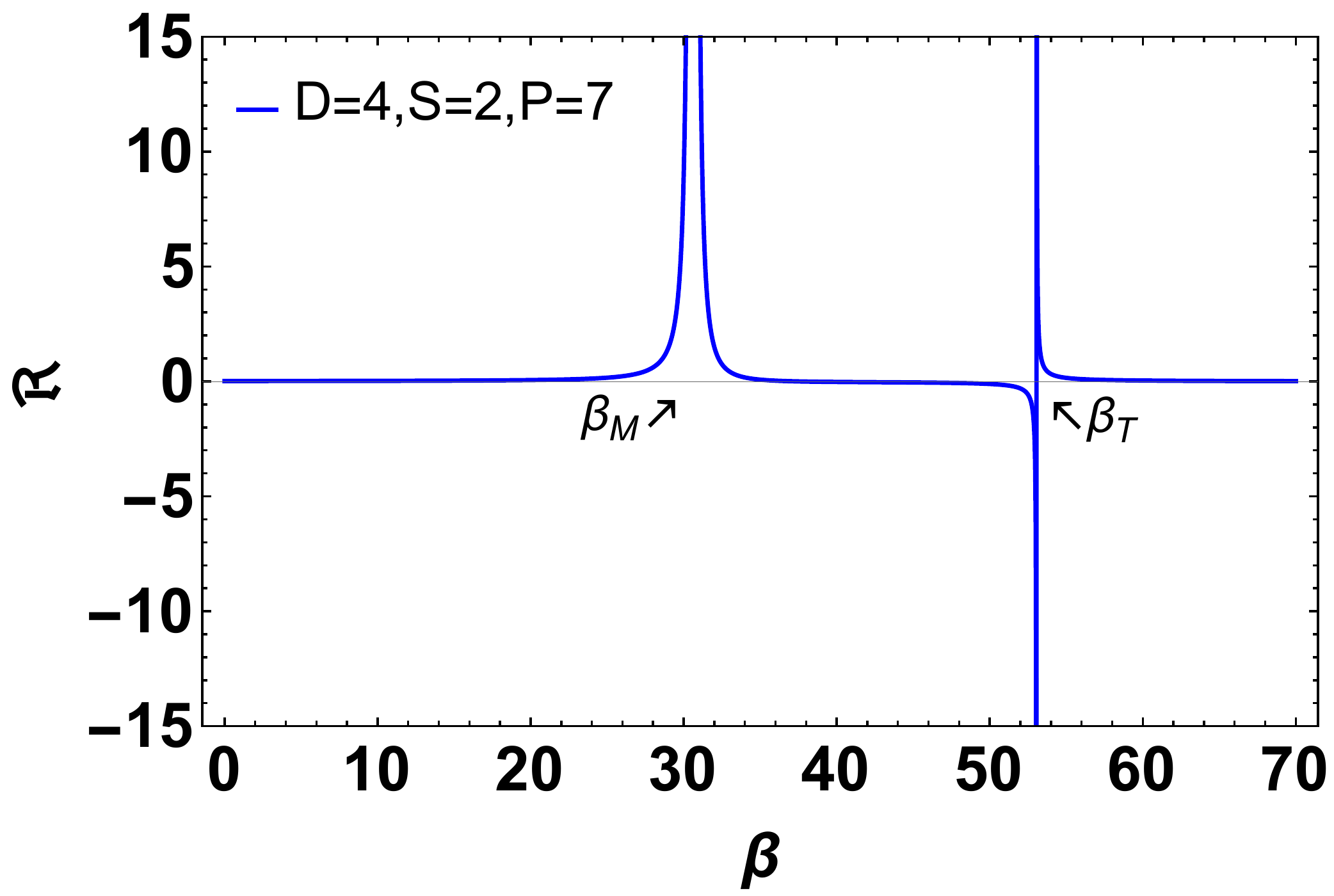}
 		\caption{}
 		\label{fig:9-1}
 	\end{subfigure}
 	\begin{subfigure}{0.45\textwidth}\includegraphics[width=\textwidth]{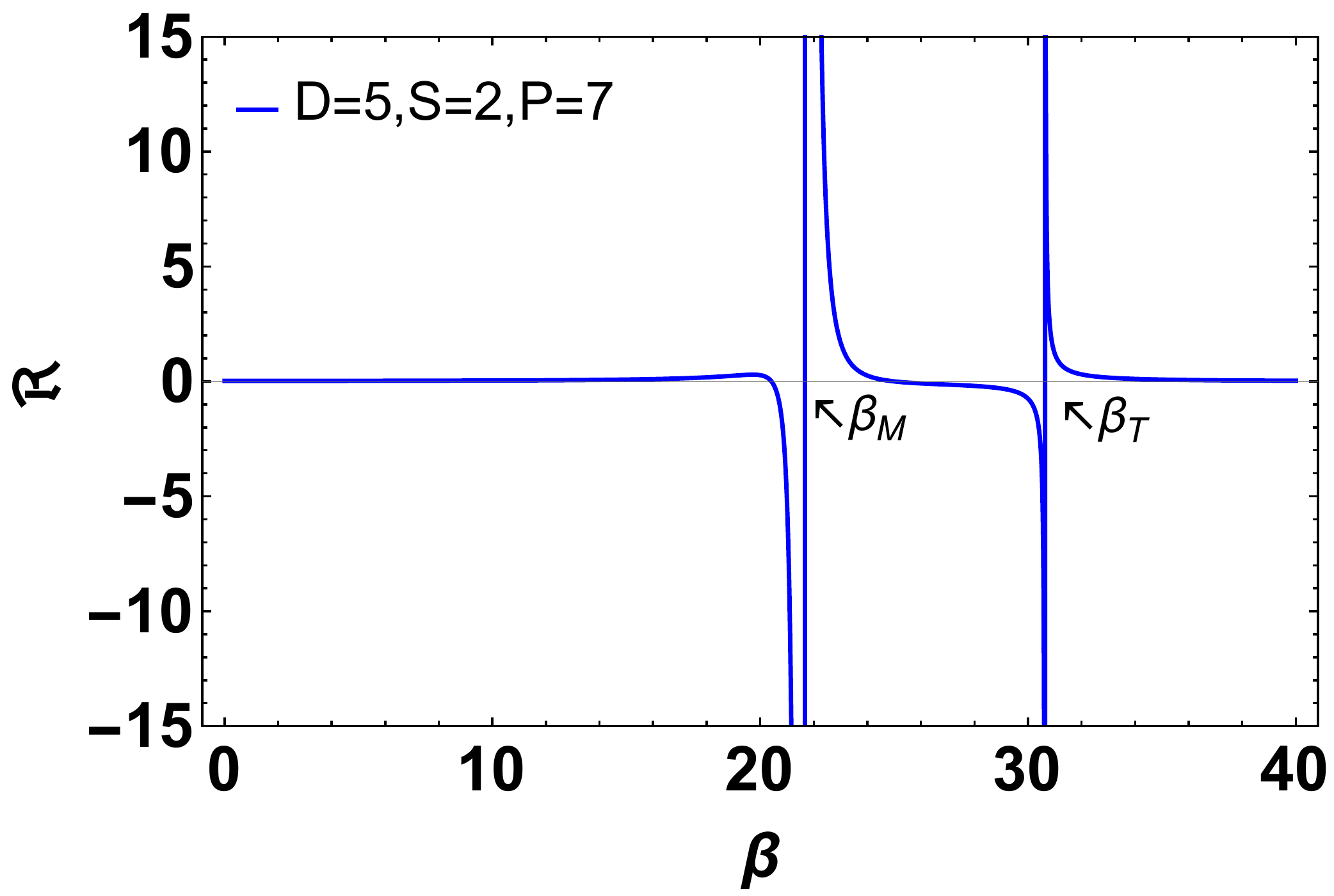}
 		\caption{}
 		\label{fig:9-2}
 	\end{subfigure}
 	\caption{Thermodynamic curvature of extended thermodynamic phase space for a black hole with momentum relaxation as a function of momentum relaxation parameter for fixed values of entropy ($S=2$) and pressure ($P= 7$) in $D= 4$ (\ref{fig:9-1}) and $D= 5$ (\ref{fig:9-2}).}
 	\label{fig:9}
 \end{figure}
 
 Similar to  the previous section, we can work out the thermodynamic curvature as a function of temperature 
\begin{equation}\label{R3TT}
\mathcal{R}= \frac{d_0P^3 S^{\frac{3}{D-2}}+d_1 P^2 S^{\frac{2}{D-2}}T+d_2 P S^{\frac{1}{D-2}} T^2+d_3T^3}{( D-2)^8 (D-1)^2 S^2\,\, (T -  T_{min})^3 \,\,T},
\end{equation}
where, the coefficients $d_0$, $d_1$, $d_2$ and $d_3$ are given by
\begin{subequations}
	\begin{align}
	d_0=2^{-\frac{12 (2 + D)}{ D-2}} ( D-3) \biggl(2^{\frac{27 D}{ D-2}} (1 + D^2) + 2^{20 + \frac{54}{ D-2}} \Bigl(-168 + D \bigl(60 + D ( 5 D-158)\bigr)\Bigr)\biggr),  
	\end{align}
	\begin{align} 
d_1=- 2^{-12 - \frac{50}{ D-2}} ( D-3) \Biggl(2^{\frac{27 D}{ D-2}} ( D-1) D^2 + 2^{17 + \frac{54}{ D-2}} (2 + 3 D) \biggl(-56 + D \Bigl(292 + D \bigl(-202 + D (-37 + 4 D)\bigr)\Bigr)\biggr)\Biggr),  
	\end{align}
	\begin{align} 
d_2=&2^{-\frac{12 (2 + D)}{ D-2}} ( D-3)\\
\times&\left(2^{\frac{25 D}{ D-2}} D^2 + 2^{15 + \frac{50}{ D-2}} \left(544 + D \left( D \Biggl(992 + D \biggl( D \Bigl(370 + D \bigl( D ( 2 D-11)-27 \bigr)\Bigr)-1240 \biggr)\Biggr)-1648 \right)\right)\right)\nonumber \,,  
	\end{align}
	\begin{align}
	d_3=-18 ( D-3) ( D-2)^7\,.
	\end{align}
\end{subequations}
Using Eq. (\ref{R3TT}), we recognize that $T=0$ and $T=T_{min}$ are two singular points of the thermodynamic curvature.
Similar to Eqs. (\ref{R3D4}) and (\ref{R3D5}), we obtain the thermodynamic curvature in four and five dimensional spacetime
\begin{equation}\label{R3D4beta}
\mathcal{R}_{D=4}=- \frac{120 P^2 S - 78 P S^{1/2} T + 9 T^2}{T (8 P S^{3/2} - 3 S T)^2},
\end{equation}
\begin{equation}\label{R3D5beta}
\mathcal{R}_{D=5}=- \frac{2560 P^3 S - 5088 2^{1/3} P^2 S^{2/3} T + 2808 2^{2/3} P S^{1/3} T^2 - 729 T^3}{36 S^2 (2 2^{2/3} P S^{1/3} - 3 T)^3 T}.
\end{equation}
Thermodynamic curvature as a function of the temperature for fixed values of the entropy and pressure has depicted in Fig. (\ref{fig:10}). Singular behaviour of the thermodynamic curvature at $T=0$ and $T=T_{min}$ is obvious. 
 \begin{figure}[h]
	\begin{subfigure}{0.45\textwidth}\includegraphics[width=\textwidth]{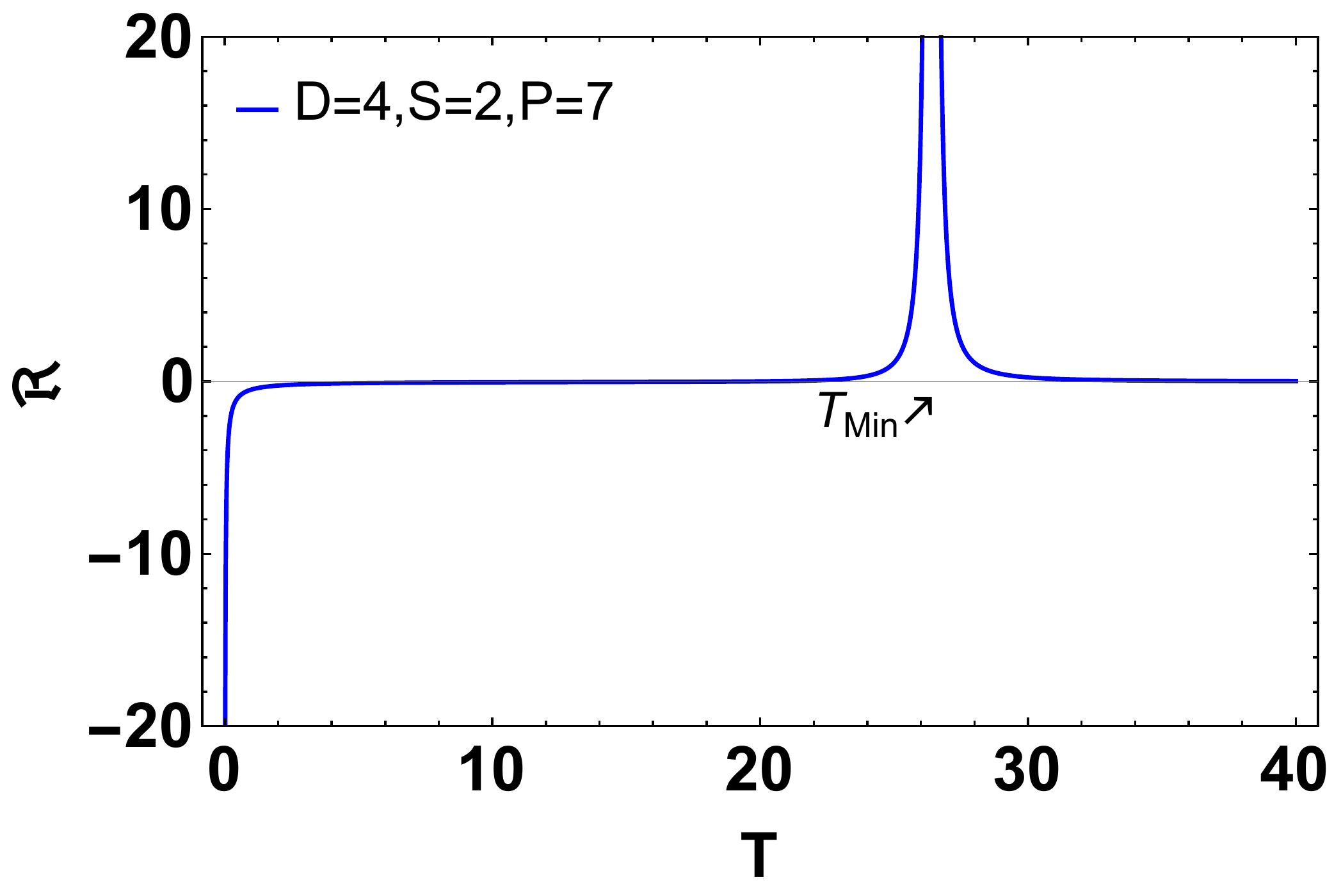}
		\caption{}
		\label{fig:10-1}
	\end{subfigure}
	\begin{subfigure}{0.45\textwidth}\includegraphics[width=\textwidth]{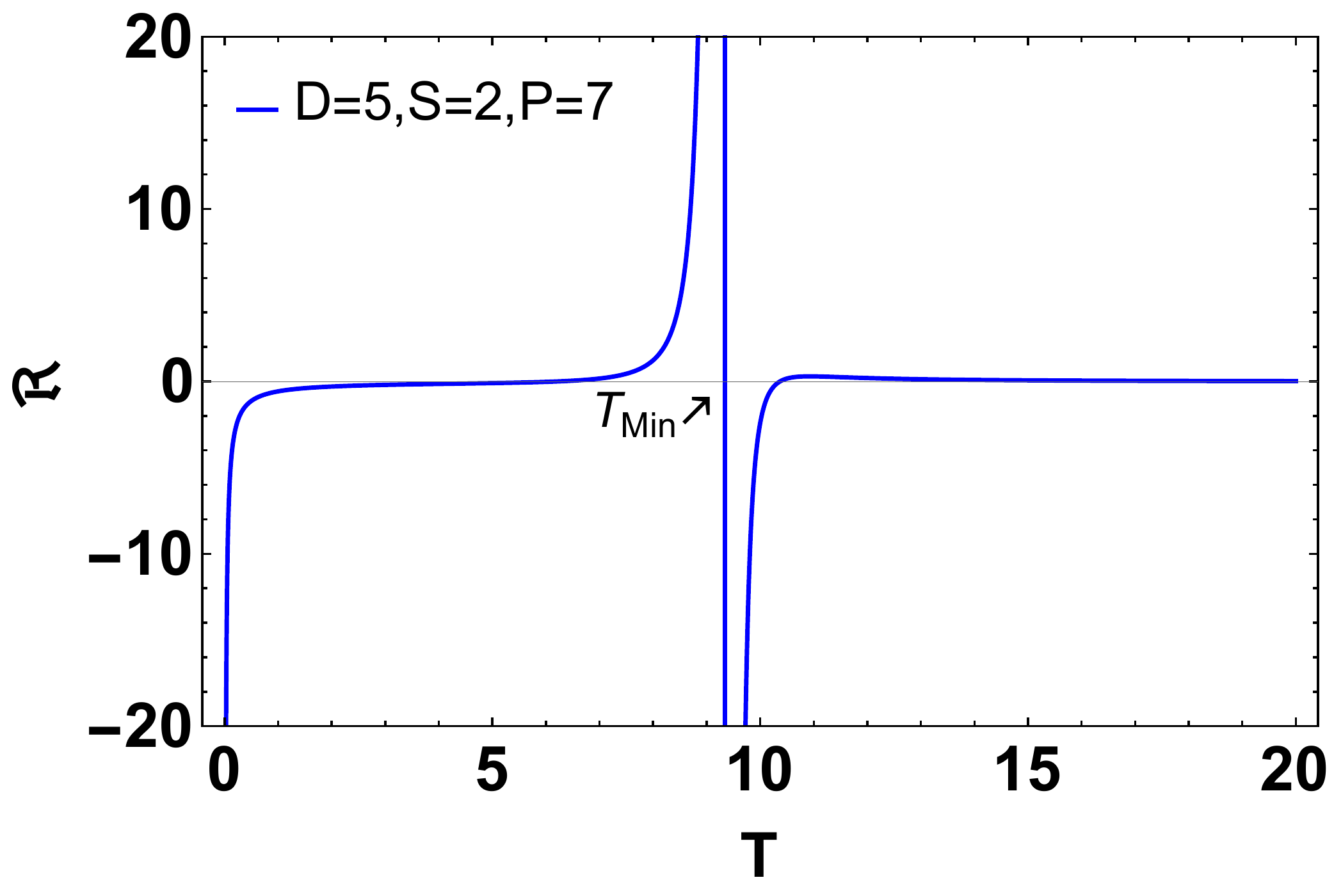}
		\caption{}
		\label{fig:10-2}
	\end{subfigure}
	\caption{Thermodynamic curvature of extended thermodynamic phase space for a black hole with momentum relaxation as a function of temperature for fixed values of entropy ($S=2$) and pressure ($P= 7$) in $D= 4$ (\ref{fig:10-1}) and $D= 5$ (\ref{fig:10-2}).}
	\label{fig:10}
\end{figure}
We consider the power-law behaviour of the thermodynamic curvature in the vicinity of $T=T_{min}$. 
For the four dimensional spacetime, we obtain that
\begin{equation}\label{RD4Tminc}
\mathcal{R}_{D=4}\propto (T -  T_{min})^{-2},
\end{equation}
while for the five dimensional spacetime, we show that
\begin{equation}\label{RD5Tim}
\mathcal{R}_{D=5}\propto -(T - T_{min})^{-3} \,.
\end{equation}
It seems that the general behaviour of the thermodynamic curvature near the critical temperature is expressed by a universal exponent defined by, $\mathcal{R}=(T-T_{c})^{-\omega}$. It has been shown that the exponent $\omega$ is related to the exponent of the heat capacity near the phase transition point. For the black holes considered in this paper, we have proved that  exponent in the thermodynamic curvature is $\omega=3$ except for a four dimensional black hole with momentum relaxation as a fluctuating thermodynamic parameter which is $\omega=2$. Therefore, we argue that the fluctuation of momentum relaxation might change the universality class of the black hole as a thermodynamic system. 
\subsection{Complexity with momentum relaxation }
Similar to the previous section, we can obtain the complexity based on the (CA) \cite{yekta2021holographic} and (CV) conjectures for the black holes with momentum relaxation. In the former case, we obtain the complexity as a function of temperature 
\begin{equation}\label{CAbeta}
\dot{\mathcal{C}}_A=2 M=\frac{2 (D-2)  S}{D-3}\,(T-T_{min})\,.
\end{equation}
It is obvious that at $T=T_{min}$, the complexity vanishes, $\dot{\mathcal{C}}_A=0$. In fact, for the non-physical range $T<T_{min}$, mass is negative and the complexity vanishes \cite{yekta2021holographic}. We showed that the thermodynamic curvature is singular at this point. 

We would like to study the late time behaviour of the holographic complexity using the (CV) conjecture. The metric (\ref{eq-metricBeta}) in Eddington-Finkelstein coordinate, becomes
\begin{equation}\label{eq:vmetricBH}
d s^{2}=-f(r) d v^{2}+2\,dv\,dr+r^2dx^a dx^a\,,
\end{equation}
where used the coordinate $v=t + r_*(r)$ with $r_*(r)= -\int^\infty_r \frac{dr'}{f(r')}$ . In \cite{carmi2017time} the final result for the rate of growth of the complexity in the neutral $\mathrm{AdS}$ black holes becomes
\begin{align}
\lim_{\tau\rightarrow\infty} \frac{d\mathcal{C}_V}{d \tau}= \frac{1}{G_N R} W(\tilde r_{min}) = \frac{1}{G_N R} \sqrt{-f(\tilde r_{min})}\, \tilde r_{min}^{D-2}\,.
\label{const_dv/dt}
\end{align}
 where $r_{min}$ is the minimum distance inside the future horizon and the $\tilde r_{min}$ is obtained solving 
 \begin{align}
 0=W'(\tilde r_{min})
 = (D-2) \tilde r_{min}^{D-3} \sqrt{-f(\tilde r_{min})} - \frac{\tilde r_{min}^{D-2} f'(\tilde r_{min})}{2\sqrt{-f(\tilde r_{min})}} \, .
 \label{r_m_eq}
 \end{align}
By replacing the $f(\tilde r_{min})$ from Eq. (\ref{fr}) in the above equations we expand Eq. (\ref{r_m_eq}) in the limit  $r_h \gg L$ to  find the leading corrections to $\tilde r_{min}$
\begin{equation}\label{rmin}
\tilde r_{min}=\frac{r_h }{2^{\frac{1}{D-1}}}\Bigl(1 + \frac{\bigl(1 + 2^{\frac{2}{ D-1}} ( D-2) -  D\bigr) L^2  \beta^2}{2 ( D-3) ( D-1)^2 rh^2} + \frac{\bigl(2^{\frac{4}{ D-1}} (D-4) (D-2) + 2^{1 + \frac{2}{ D-1}} ( D-1) -  (D-1)^2\bigr) ( D-2) L^4 \beta^4}{8 ( D-3)^2 ( D-1)^4 rh^4}+...\Bigr)\, .
\end{equation}
We obtain the rate of growth of the complexity of black holes with the momentum relation
	\begin{equation}\label{CV1beta}
	\dot{\mathcal{C}}_V=
	\frac{8 \pi }{  D-2}(M-M_{min}) \bigg(\frac{  D-3}{ D-1} +  \frac{\bigl( 4 D-8- 4^{\frac{1}{D-1}} ( D-1) \bigr) L^2 \beta^2}{4 ( D-1)^2 r_h^2} + \frac{\bigl(16 ( D-2) -  2^{3 + \frac{2}{ D-1}} ( D-1) + 2^{\frac{4}{ D-1}} ( 3 D-5)\bigr) L^4 \beta^4}{32 ( D-1)^3 r_h^4}+...\bigg),
	\end{equation}
where $M_{min}$ corresponds to $T=0$ and given by
\begin{equation}\label{MminBeta}
M_{min}=- \frac{( D-2)  }{8 \pi ( D-3)  } \frac{ r_h^{ D-1} }{ L^2 }.
\end{equation}
One can show that always the following relation is established:
\begin{eqnarray}
\dot{\mathcal{C}}_V \geq
\frac{8 \pi }{  D-2} \frac{  D-3}{ D-1} (M-M_{min})\,.
\end{eqnarray}
The volume equivalent to the complexity in $M=M_{min}$ equals zero.
$\dot{\mathcal{C}}_A$ and $\dot{\mathcal{C}}_V$ as a function of temperature for fixed values of $S$ and $P$ have been depicted in Fig. (\ref{fig:8}). 
\begin{figure}[h]
	\begin{subfigure}{0.45\textwidth}\includegraphics[width=\textwidth]{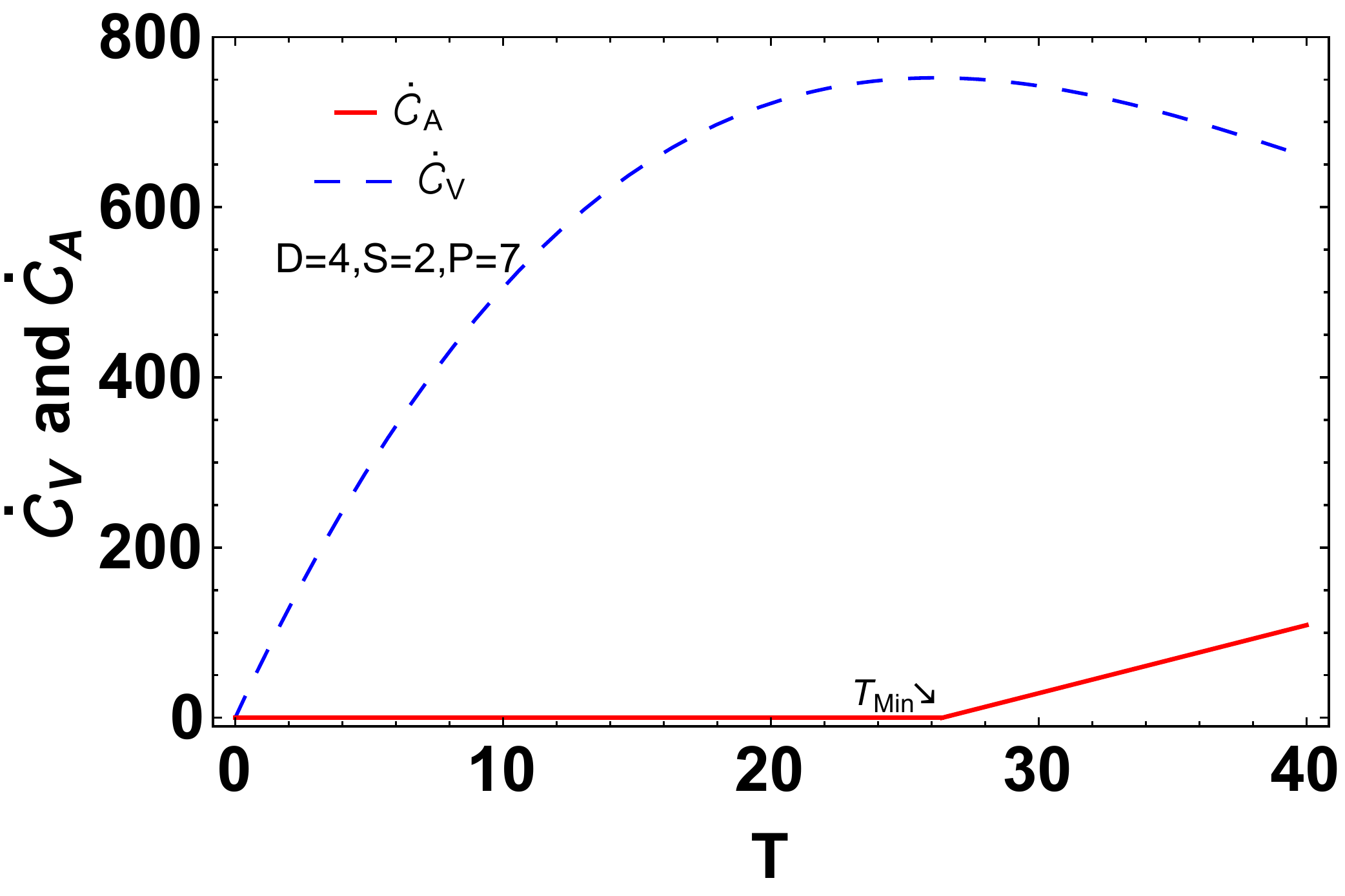}
		\caption{}
		\label{fig:8-1}
	\end{subfigure}
	\begin{subfigure}{0.45\textwidth}\includegraphics[width=\textwidth]{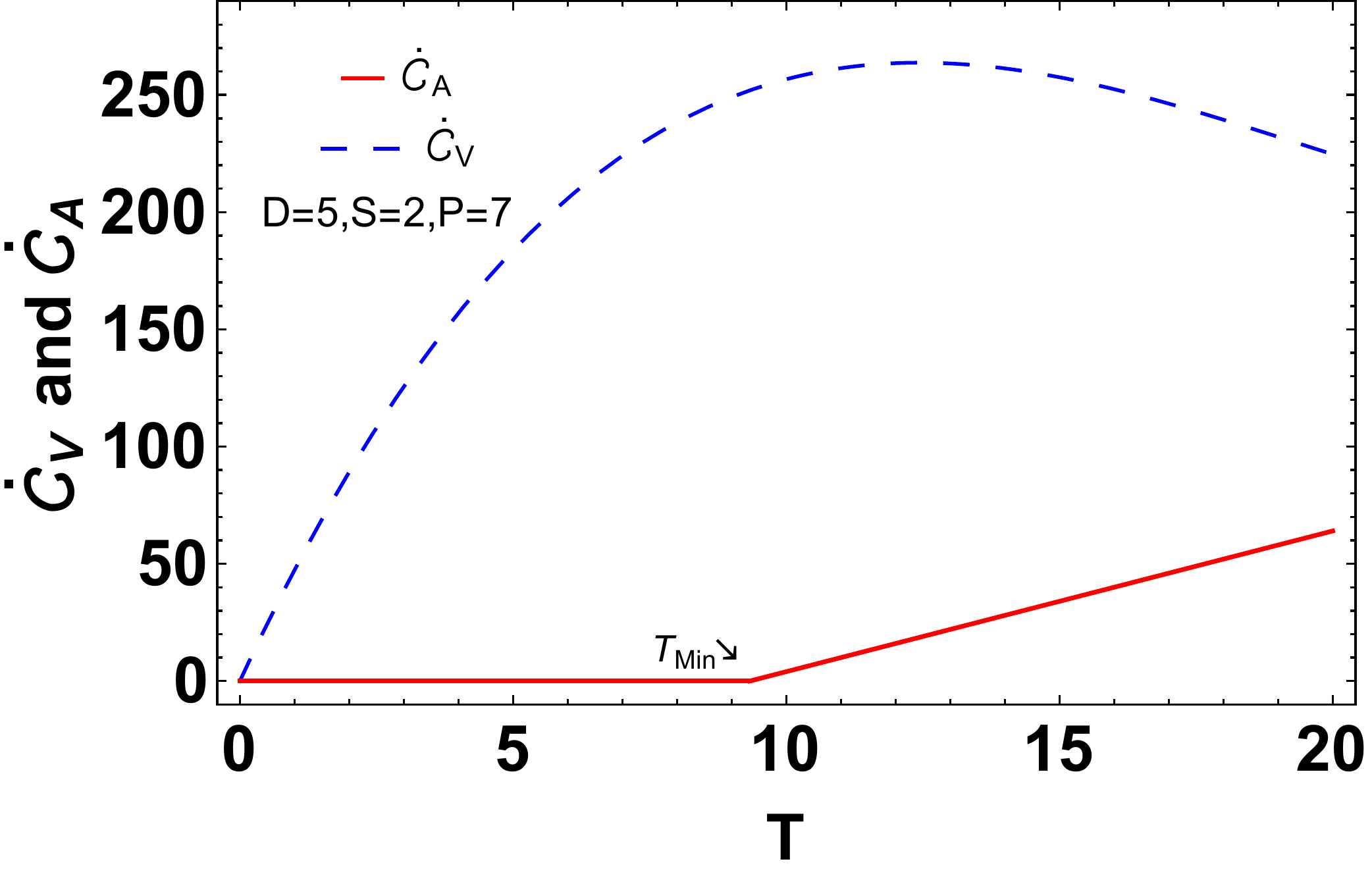}
		\caption{}
		\label{fig:8-2}
	\end{subfigure}
	\caption{The Complexity (equivalent to action; red solid line) and (equivalent to volume; blue dashed line)  as a function of temperature for fixed values of entropy $S=2$ and pressure $P=7$ in $D=4$ (\ref{fig:8-1}) and $D=5$ (\ref{fig:8-2}).}
	\label{fig:8}
\end{figure}

Now, we focus on the behaviour of a thermodynamic response function at singular points of thermodynamic curvatures which are correspond to the zero points of complexity. Heat capacity at constant pressure and momentum relaxation temperature is defined
\beq\label{CPbeta1}
{{C}_{P \beta}}={{\left( \frac{\partial M}{\partial T} \right)}_{P,\beta}}=\frac{{{\left( \frac{\partial M}{\partial S} \right)}_{P,\beta}}}{ {{\left( \frac{\partial T}{\partial S} \right)}_{P,\beta}}}\,.
\eeq
We use the following relation to evaluate the partial differentiation
\begin{equation}
{{\left( \frac{\partial f}{\partial g} \right)}_{h,k}}=\frac{{{\left\{ f,h,k \right\}}_{{{q}_{1}},{{q}_{2}},{{q}_{3}}}}}{{{\left\{ g,h,k \right\}}_{{{q}_{1}},{{q}_{2}},{{q}_{3}}}}}\,,
\label{q1q2q3}
\end{equation}
where the Nambu bracket is defined
\begin{equation}
{{\left\{ f,h,k \right\}}_{{{q}_{1}},{{q}_{2}},{{q}_{3}}}}=\sum\limits_{ijk=1}^{3}{{{\varepsilon }_{ijk}}\frac{\partial f}{\partial {{q}_{i}}}}\frac{\partial h}{\partial {{q}_{j}}}\frac{\partial k}{\partial {{q}_{k}}}\, ,
\label{N31}
\end{equation}
and $\varepsilon_{ijk} $ is the Levi-Civita symbol. We obtain that
\begin{eqnarray}\label{CPbeta2}
C_{P \beta}&=&\frac{( D-2) S \bigl(2^{5 + \frac{4}{D-2}} P \pi S^{\frac{2}{ D-2}} - ( D-2)  \beta^2\bigr)}{2^{5 + \frac{4}{ D-2}} P \pi S^{\frac{2}{ D-2}} + ( D-2)  \beta^2}=-(D-2)\frac{ (  \beta^2-\beta^2_T)}{ \beta^2+\beta^2_T} S\,.
\end{eqnarray}

The sign of heat capacity is changed at $\beta=\beta_T$. We know that $\beta_{T}$ corresponds to $T=0$. Therefore, the heat capacity vanishes at zero temperature. We can extract the heat capacity as an explicit function of temperature as follows
\begin{eqnarray}\label{CPbetaTT}
C_{P \beta}&=&\frac{( D-2)^2 S T}{2^{3 + \frac{2}{ D-2}} P S^{\frac{1}{-2 + D}}- ( D-2) T}\,,
\end{eqnarray}
which shows that at $T=0$, the heat capacity vanishes.
\end{widetext}
\section{Conclusion}\label{4}
We considered the thermodynamic geometry, action complexity, and volume complexity of two black holes. First, we constructed the thermodynamic geometry of a hyperbolic black hole in arbitrary dimensions. For a two dimensional thermodynamic parameters space, we obtained the thermodynamic curvature and investigated its behaviour near the critical point. Similar evaluations were performed on the black hole with momentum relaxation. Of course, we consider the momentum relaxation parameter as a constant or fluctuating thermodynamic parameter in two different cases.

Using the singular points of the thermodynamic curvature and the sign change points of a response function, namely the heat capacity, we recognized the critical points. We obtain the critical exponent of thermodynamic curvature in the vicinity of the phase transition temperature.

We showed that the singular points of thermodynamic curvature correspond to the zero points of the action and the volume complexities. At $T=0$ the volume complexity vanishes and the thermodynamic curvature is singular for two considered black holes in all dimensions. The action complexity vanishes at $T=T_{min}$, while the thermodynamic curvature is singular at this point and behaves such as power law function in the vicinity of the critical temperature.

It is well known that for a magnetic system, the magnetization as the order parameter vanishes at the paramagnetic phase and grows up at the ferromagnetic phase. The behaviour of the action complexity with respect to the temperature is interesting and similar to magnetization of the magnetic systems. In fact, one can define two different phases using the action complexity. For $T\le T_{min}$, the action complexity vanishes and it grows up for $T>T_{min}$. According to the singular behaviour of thermodynamic curvature and special character of the action complexity at $T=T_{min}$, one may propose the action complexity as an order parameter of the black holes as a thermodynamic system. 


\bibliography{refs}

\end{document}